\documentclass{ieeeaccess}
\usepackage{cite}
\usepackage{amsmath,amssymb,amsfonts}
\usepackage{algorithmic}
\usepackage{mathtools}
\usepackage{graphicx}
\usepackage[caption=false]{subfig}
\usepackage{color}
\usepackage{blindtext}
\usepackage{hyperref}
\usepackage{float}
\usepackage{booktabs}
\usepackage{graphicx}
\usepackage{textcomp}
\usepackage{soul} 
\usepackage{color, xcolor} 
\definecolor{OliveGreen}{rgb}{0,0.6,0}
\DeclarePairedDelimiter\abs{\lvert}{\rvert}
\DeclarePairedDelimiter\norm{\lVert}{\rVert}
\makeatletter
\let\oldabs\abs
\def\abs{\@ifstar{\oldabs}{\oldabs*}}
\let\oldnorm\norm
\def\norm{\@ifstar{\oldnorm}{\oldnorm*}}
\makeatother

\begin{document}
\history{Date of publication xxxx 00, 0000, date of current version xxxx 00, 0000.}
\doi{10.1109/ACCESS.2023.0322000}

\title{Sparse Variable Selection on High Dimensional Heterogeneous Data with Tree Structured Responses}
\author{\uppercase{Hui Liu}\authorrefmark{1,$\dag$}, \uppercase{Xiang Liu}\authorrefmark{2,$\dag$}, {Jing Diao}\authorrefmark{3} Wenting Ye\authorrefmark{4}, Xueling Liu\authorrefmark{5}, and Dehui Wei\authorrefmark{6}}

\address[1]{Jiangxi Normal University, Jiangxi, 330022, PR China (e-mail: huiliu.jxnu@gmail.com)}
\address[2]{School of Computing, National University of Singapore, 119077 Singapore (e-mail: liuxiang@comp.nus.edu.sg)}
\address[3]{Department of Preventive Dentistry, Peking University School and Hospital of Stomatology, Beijing, 100081, PR China (e-mail: diaoj@pku.edu.cn)}
\address[4]{School of Computer Science, Carnegie Mellon University, Pittsburgh, PA 15213, USA (e-mail: wye2@alumni.cmu.edu)}
\address[5]{Yong Loo Lin School of Medicine, National University of Singapore, 117597 Singapore (e-mail: lxlxueling@gmail.com)}
\address[6]{State Key Laboratory of Networking and Switching Technology, BUPT, Beijing, 100876, China. (e-mail: dehuiwei@bupt.edu.cn)}
\address[$\dag$]{Equal Contribution. The order of authors' names is alphabetical.}

\markboth
{Author \headeretal: Preparation of Papers for IEEE TRANSACTIONS and JOURNALS}
{Author \headeretal: Preparation of Papers for IEEE TRANSACTIONS and JOURNALS}

\corresp{Corresponding author: Hui Liu (e-mail: huiliu.jxnu@gmail.com).}

\begin{abstract}
We consider the problem of sparse variable selection on high dimension heterogeneous data sets, which has been taking on renewed interest recently due to the growth of biological and medical data sets with complex, non-i.i.d. structures and huge quantities of response variables. The heterogeneity is likely to confound the association between explanatory variables and responses, resulting in enormous false discoveries when Lasso or its variants are na\"ively applied. Therefore, developing effective confounder correction methods is a growing heat point among researchers. However, ordinarily employing recent confounder correction methods will result in undesirable performance due to the ignorance of the convoluted interdependency among response variables. To fully improve current variable selection methods, we introduce a model, the tree-guided sparse linear mixed model, that can utilize the dependency information from multiple responses to explore how specifically clusters are and select the active variables from heterogeneous data. Through extensive experiments on synthetic and real data sets, we show that our proposed model outperforms the existing methods and achieves the highest ROC area.
\end{abstract}


\titlepgskip=-21pt

\maketitle

\section{INTRODUCTION}
\label{sec:introduction}
\PARstart{V}{a}riable selection is one of the central tasks in statistics and has been studied for decades \cite{cai2010unsupervised,du2015unsupervised}. Modern machine learning problems, especially biological or medical applications often seek solutions in the existing statistical approaches. Lasso \cite{tibshirani1996regression} is an example of those widely adopted methods in a variety of areas for sparse variable selection tasks. However, the increasing volume of data sets often requires the data to be collected from multiple batches and then integrated together. This procedure is particularly harmful to the biological \cite{he2011variable} and medical \cite{chen2013variable,zhou2013feafiner} data sets, which are sensitive to the data sources, like populations, hospitals or even experimental devices. This sensitivity results in the heterogeneity, therefore, breaks one of the most fundamental assumptions (i.i.d. assumption) that most statistical machine learning methods make. More importantly, due to the expensiveness of biological and medical data, different batches of data are gathered for different purposes from distinctly different sources, such as samples of the control group are mostly collected from volunteers from several different undeveloped regions. Consequently, the heterogeneity often induces confounding factors between explanatory variables and response variables, resulting in numerous false positive selected variables when classical variable selection techniques are applied \cite{astle2009population}.

To deepen understanding of the challenge that heterogeneity is introduced to biological or medical data sets and define the problem, consider that we have data samples in the format of $(X, Z, Y)$, where $X$ stands for the explanatory variables, $Y$ stands for the responses, $Z$ stands for the indicator of the data source. The dependency between $X$ and $Y$ is the premise of any variable selection tasks \cite{rakitsch2013lasso}, and the dependency between $X$ and $Z$ is induced through heterogeneity \cite{wang2017variable,wang2017multiplex}. The data collection procedure we mentioned brings the dependency between $Z$ and $Y$. In the real world, this problem may be even intractable, for the origin of different samples is lost either through data compression or experimental necessity in most cases. Nowadays genetic association studies are rarely aware of the origin of the samples listed. $Z$ becomes the confounding factor between $X$ and $Y$ \cite{henderson1975best,Wang2019,Dinga2020,Hatoum2020}. One challenge of the heterogeneous data variable selection problem is to mitigate the confounding effects brought by $Z$. 


Aside from challenges above, many of the real world biological and medical data sets are collected along with multiple response variables. These responses are often more closely related and could share common relevant covariates than others and then form the tree or other kinds of structures \cite{chen2010graph,kim2012tree,ye2017sparse,wang2019graph}. For instances, in genetic association analysis, which aims to select the single-nucleotide polymorphism (explanatory variables) that could affect the phenotype (response variables), the genes in the same pathway pretend to share the common set of relevant explanatory variables than other genes. 

Thus, to improve the performance of the variable selection, incorporating the complex correlation structure in the responses is under our consideration. In this paper, we extend the recent solutions of sparse linear mixed model \cite{rakitsch2013lasso,wang2017variable} that can correct confounding factors and perform variable selection simultaneously further to account the relatedness between different responses. We propose the \textbf{t}ree-\textbf{g}uided \textbf{s}parse \textbf{l}inear \textbf{m}ixed \textbf{m}odel, namely TgSLMM, to correct the confounder and incorporate the relatedness among response variables simultaneously. With TgSLMM, we are capable to improve the performance of the variable selection when considering the statistical criterion, incorporating the complex tree-based correlation structure in the traits under our consideration. Eventually, we examine our model through plenty of repeated experiments and show that our method is superior to other existing approaches and able to discover the real genome association in the real data set.

\section{RELATED WORK}
\label{sec:related}

Recent years have witnessed the great advances in the variable selection area. The most classical approach is $\ell_{1}$-norm regularization (i.e. \emph{Lasso} regression \cite{tibshirani1996regression}). Further, studies have extended the model capability by introducing various regularizers \cite{chen2010graph}. Examples including the Smoothly Clipped Absolute Deviation (SCAD) \cite{fan2001variable}, the Local Linear Approximation (LLA) \cite{zou2008one}, the Minimax Concave Penalty (MCP) \cite{zhang2010nearly}, and the Precision Lasso \cite{wang2018precision} have been introduced since then, which all overcome a variety of limitations of Lasso \cite{fan2001variable}. Some other variable selection methods like \cite{kolda2009tensor} ignore underlying multidimensional structure, leading to severe small dataset problems. \cite{tan2012logistic} imposes a rank constraint into $\ell_1$ regularization to factor matrices and promotes sparsity in variable selection, which hurts the interpretability. The liability-threshold mixed linear model overcomes the limitation of Linear Mixed Model (LMM) in case-control ascertainment \cite{hayeck2015mixed}. \cite{liu2016consensus} proposed a unsupervised variable selection method. But both of them cannot apply to high dimensional data with heterogeneity.

Besides these, in the non-i.i,d setting, confounders could raise a challenge in variable selection when the data set is originated from different sources. Corresponding solutions have been studied for decades. Principal components analysis (PCA) \cite{patterson2006population,price2006principal} and linear mixed model \cite{goddard2009genomic,kang2010variance} are two popular and efficient approaches to alleviate the confounding effect. The latter provides a more fine-grained way to model the population structure and won its prominence in the animal breeding literature, where it was used to reveal the underlying kinship and family structure \cite{henderson1975best,wang2022trade}. Many extensions have been developed, however, these measures such as LMM-Select \cite{listgarten2012improved} LMM-BOLT \cite{loh2015efficient} and Liability-threshold mixed linear model (LTMLM) \cite{loh2015efficient} along with other algorithms \cite{lippert2011fast,segura2012efficient,pirinen2013efficient} only rely on univariate testing to select the variable once uncovering the confounding factor. Attempts have been made to propose multi-variable testing model \cite{bondell2010joint,fan2012variable,rakitsch2013lasso,wang2017variable} these days, but their performances fall short while tackling with the challenge that takes the relatedness between responses into account. \cite{schwartzman2019simple,pazokitoroudi2020efficient,wang2020discovering,jiang2021generalized,wu2022fast,wang2022gene,st2023efficient} are proposed to identify significant associations, which is to be contrasted
with the related problem of estimating heritability. However, they also lack accounting
for the relatedness between different traits \cite{kalantzis2022methods}. \cite{hou2023genetic} helps improve association methods for kinship estimation, but it could not construct the convoluted phenotypic architecture in a dataset originated from different populations in the real world like \cite{kim2012tree}. The challenges show the desire to have a method, which requires no prior knowledge of the individual relationship and is capable of uncover the structured pattern in a way that is properly calibrated to the degrees of traits’ relatedness.

\section{TREE-GUIDED SPARSE LINEAR MIXED MODEL}
\label{sec:model}
Throughout this paper, $X$ denotes the $n \times p$ matrix for explanatory variables for individuals, $Y$ denotes the $n \times k $ matrix for response variables, and $\beta$ denotes the $p \times k$ matrix for effect sizes. We use subscripts to denote rows and superscripts to denote columns, for example, $\beta_k$ and $\beta^k$ are the $k$-th column and $k$-th row of $\beta$ respectively.

In this section, we begin by examining the sparse linear mixed model. Next, we demonstrate how we leverage the technique to uncover relationships between traits. Finally, we transform this approach into a regression problem and employ efficient methods to address associated challenges.

\subsection{\textbf{Sparse Linear Mixed Model}}
\label{sec:3.1}
The linear mixed model (LMM) is an extension of the standard linear regression model that explicitly describes the relationship between response variables and explanatory variables incorporating an extra random term to account for confounding factors. To introduce the sparse linear mixed model, we briefly revisit the classical linear mixed model as Equation~\ref{LMM}: 

\begin{equation}
\label{LMM}
Y = X\beta +Zu +\epsilon 
\end{equation}

where $Z$ is an $n \times t$ matrix for the random effect. $u$ is the confounding influences with implicitly identity correlation information, $\epsilon$ denotes observed noise and they both follow the independent Gaussian distribution with the zero means. Intuitively, $Zu$ models the covariance between the observations $y_i$. Assuming that $\epsilon\sim\,\mathcal{N}(\,0,\,\sigma_\epsilon^2I)$, $u\sim\,\mathcal{N}(0, \sigma_g^2I)$, $K = ZZ^T$ and represents the covariance between the responses and $\sigma_g$ represents the magnitude of confounder factors, we can rewrite the formula as Equation~\ref{LMM2} to simplify mathematical derivation:
\begin{equation}
\label{LMM2}
y_k \sim  \mathcal{N}\,(X\beta_k, \sigma_g^2K + \sigma_\epsilon^2I) 
\end{equation}

Assuming the priori distribution of $\beta$ could be expressed as $e^{-\Phi(\beta)}$, we can define log likelihood function as Equation~\ref{log likelihood}:

\begin{equation} 
\label{log likelihood}
\ell(\sigma_g^2, \sigma_\epsilon^2, \beta)= e^{-\Phi(\beta)}\cdot\prod_{k=1}^K\mathcal{N}\,(y_k|X\beta_k,\,\sigma_g^2K + \sigma_\epsilon^2I\,)
\end{equation}
Based on the sparsity of $\beta$, it's reasonable to assume that $\beta$ follows Laplace shrinkage prior. Such assumptions lead to the sparse linear mixed model. However, sparse LMM fails to consider the relatedness among response variables. The defect drives us to the tree-guided sparse linear mixed model.

\subsection{Tree-Guided Sparse Linear Mixed Model}
\label{sec:3.2}
To incorporate the relatedness among responses simultaneously, we use Tree-Lasso as Equation~\ref{tree-lasso penalty}.

\begin{equation}
\label{tree-lasso penalty}
\Phi(\beta) = \lambda\sum_{j} \sum_{v\in V} w_v||\beta_j^{G_v}||_2 = \lambda\sum_{j} W_j(v_{root})
\end{equation}

where $\lambda$ is a tuning parameter that controls the amount of sparsity in the solution and $\beta_j^{G_v}$ is a vector of regression coefficients $\{\beta_j^k|k \in G_v\}$. The overlaps of groups of Tree-Lasso and the number of the trees is determined by the hierarchical clustering tree. Each node $v\in{V}$ of the $j$-th tree is associated with the group $G_v$ whose members are the response variables at the nodes of the same subtree. Each group of subtree regression coefficients $\beta_j^{G_v}$ is weighted with $w_v$, which is defined as the Equation~\ref{w_v}. In general, $h_v$ in the Equation~\ref{w_v} represents the weight for selecting relevant covariates separately for the responses associated with each child of node $v$, whereas the $1 - h_v$ represents the weight for selecting relevant covariates jointly for the responses for all of the children of node $v$, and the value of $h_v$ ranges from 0 to 1. Assuming $K$ is the number of response variables and $|V|$ is equivalent to the number of nodes in one tree, since a tree associated with $K$ responses has $2K-1$ nodes, $|V|$ appears in the tree-lasso penalty is upper-bounded by $2K$. 
\begin{equation}
\label{w_v}
\small{   
w_v=\left\{
  \begin{array}{rcl}
    (1-h_v)\prod\limits_{m\in Ancestors(v)}{h_m}& &\mbox{if v is an internal node,}\\
    \prod\limits_{m\in Ancestors(v)}{h_m}& &\mbox{if v is a leaf node.}
  \end{array}
\right. 
}
\end{equation}

To simplify the computation process of Tree-Lasso, we can calculate the separate penalty from the root of each tree iteratively as Equation~\ref{tree_iteratively}:

\begin{equation}
\label{tree_iteratively}
\small{   
W_j(v)=\left\{
  \begin{array}{lr}
    (1-h_v)\sum\limits_{c \in Children(v)}|W_j(c)|+h_v||\beta_j^{G_v}||_2, &\\         \ \mbox{if v is an internal node.} &\\
    \sum\limits_{m\in G_v}|\beta_j^{m}|, &\\
    \ \mbox{if v is a leaf node.}
  \end{array}
\right. 
}
\end{equation}


\subsection{Parameter Learning}
\label{sec:3.3}
Overall, optimizing Equation~\ref{log likelihood} with hyper-parameter \{$\Theta = \sigma_g^2,\, \sigma_\epsilon^2,\, \lambda,\, w_v$\} is a non-convex optimization problem aside with weights $\beta$. Hence, we could apply the null-model fitting method first to correct the confounding factors and then solve Tree-Lasso regression problem using smoothing proximal gradient method \cite{chen2012smoothing}.

\subsubsection{Null-model}
Due to sparsity of $\beta$, null-model fitting method by first optimizing $\sigma_g^2$, $\sigma_\epsilon^2$ while ignoring individual explanatory variables, can yield near-identical result as an exact method \cite{kang2010variance}. By using the computational trick \cite{lippert2011fast} that introduces the ratio of the random effect and the noise variance, $\delta=\sigma_\epsilon^2 / \sigma_g^2 $, we could transform the equation as Equation~\ref{log likelihood reduced}:

\begin{equation}
\label{log likelihood reduced}
\ell_{null}(\sigma_g, \delta) = e^{-\Phi(\beta)} \cdot \prod_{k=1}^K\mathcal{N}\,(y_k|X\beta_k,\,\sigma_g^2(K + \delta{I}))
\end{equation}

The genetic effects are treated as fixed effects, whereas the confounding influences are modeled as random effects. We carry out a log likelihood optimization with regard to $\delta$ and then $\sigma_g$ in closed form. 

\subsubsection{Reduction to Tree-Lasso Regression Problem} 
In general, we first compute the spectral decomposition of $K = U\textnormal{diag}(d)U^T$, where $U$ for eigenvector matrix and $\textnormal{diag}(d)$ for eigenvalue matrix. Having the yielded $\delta$, we use the $U$ to reweight the data such that the covariance matrix becomes isotropic. Assume $\tilde{Y}$ and $\tilde{X}$ are the resulting rescaled data, which can be calculated by the following equation:

\[\tilde{X} = (\textnormal{diag}(d) + \delta{I})^{-\frac{1}{2}}U^TX\]
\[\tilde{Y} = (\textnormal{diag}(d) + \delta{I})^{-\frac{1}{2}}U^TY\]

Using this transformation, the equation eventually ends up with a standard Tree-Lasso regression problem since it is free of population structure and has been alleviated the confounding factor. In the following step, we can obtain the $\widehat{\beta}^{tree}$ as Equation~\ref{standard tree lasso}:

\begin{equation}
\label{standard tree lasso}
\widehat{\beta}^{tree} = \min  \limits_{\beta}\frac{1}{2}||\tilde{Y} - \tilde{X}\beta||_F^2 + \Phi(\beta)
\end{equation}

where $||\cdot||_F$ denotes the matrix Frobenius norm, and $\Phi$ is determined by the Equation~\ref{tree-lasso penalty}, then we can easily employ the smoothing proximal gradient descent method.

 \begin{figure}
\centering
  \includegraphics[width=.465\textwidth]{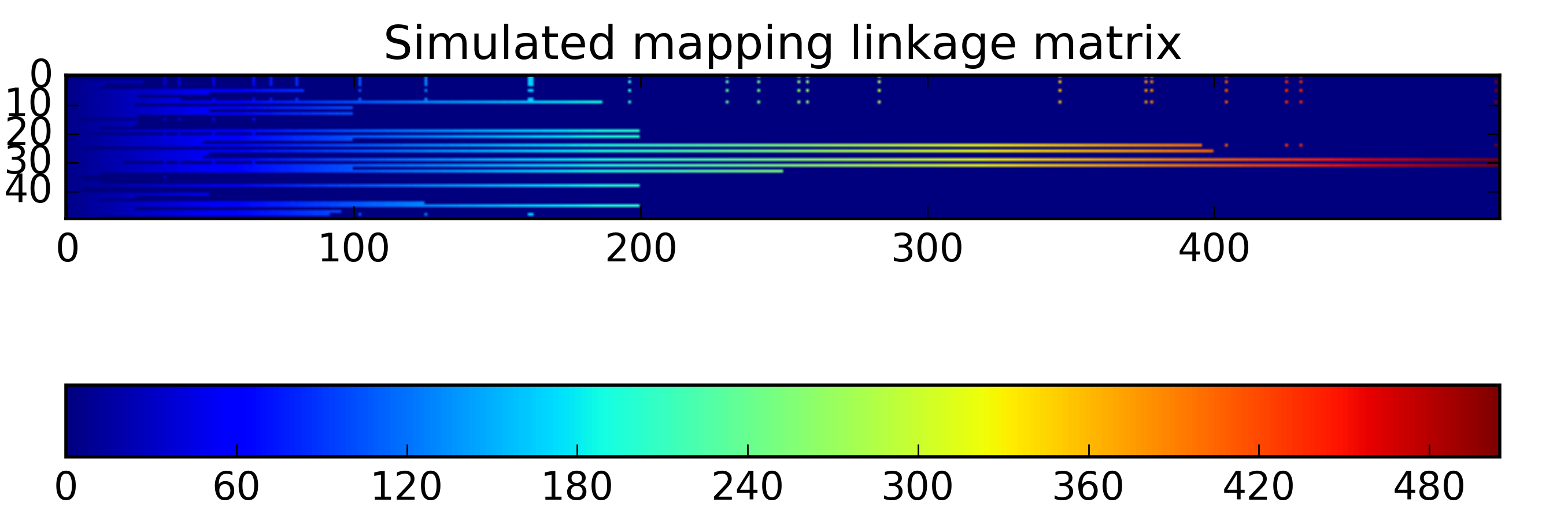}
  \caption{The simulated ground-truth $\beta$ vector. For the illustration purpose, we choose the experimental setting of $n=250$, $p =500$ and $k = 50$.}
  \label{fig:beta_cut}
\end{figure}

\begin{figure*}[t!]
\centering
\includegraphics[width=.3\textwidth]{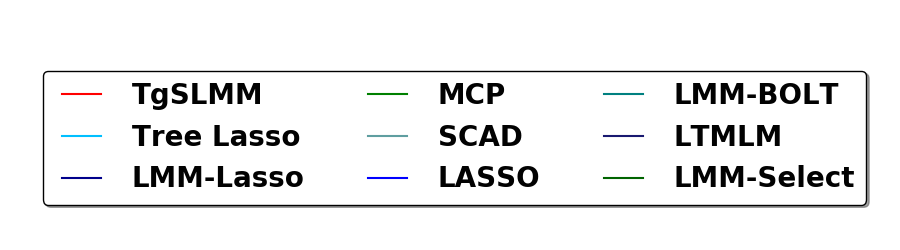}
\\
\subfloat[Different number of samples]{%
  \includegraphics[width=.475\textwidth]{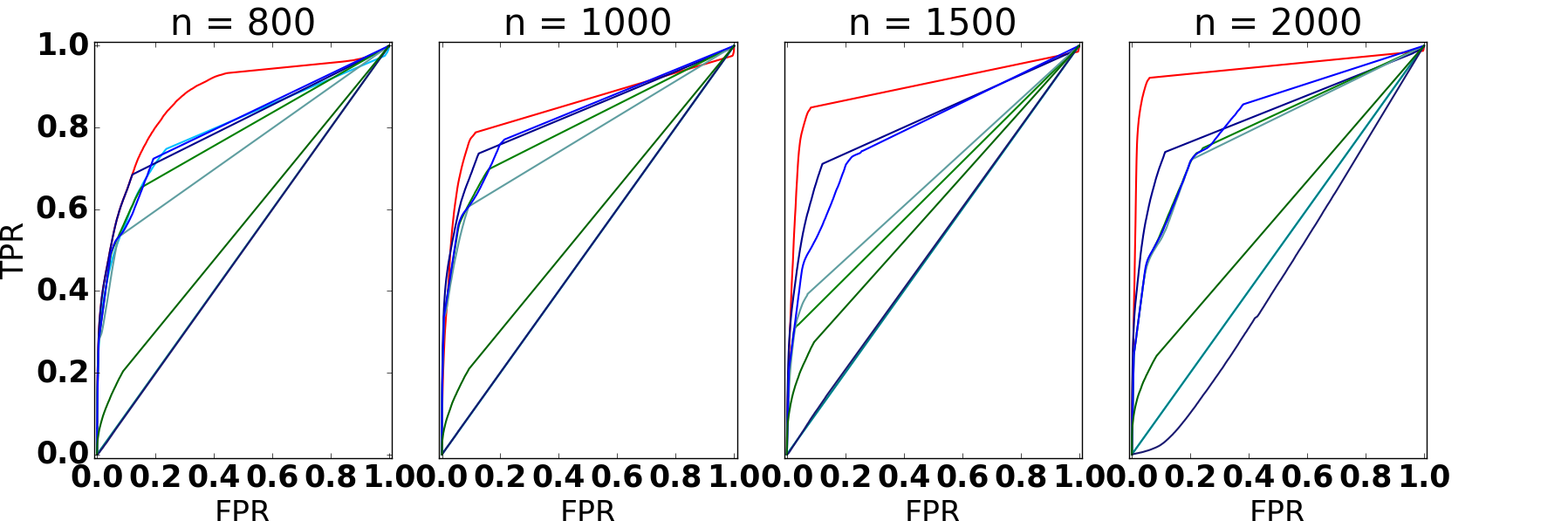}
    \label{exn}}\hfill
\subfloat[Different number of explanatory variables]{%
  \includegraphics[width=.475\textwidth]{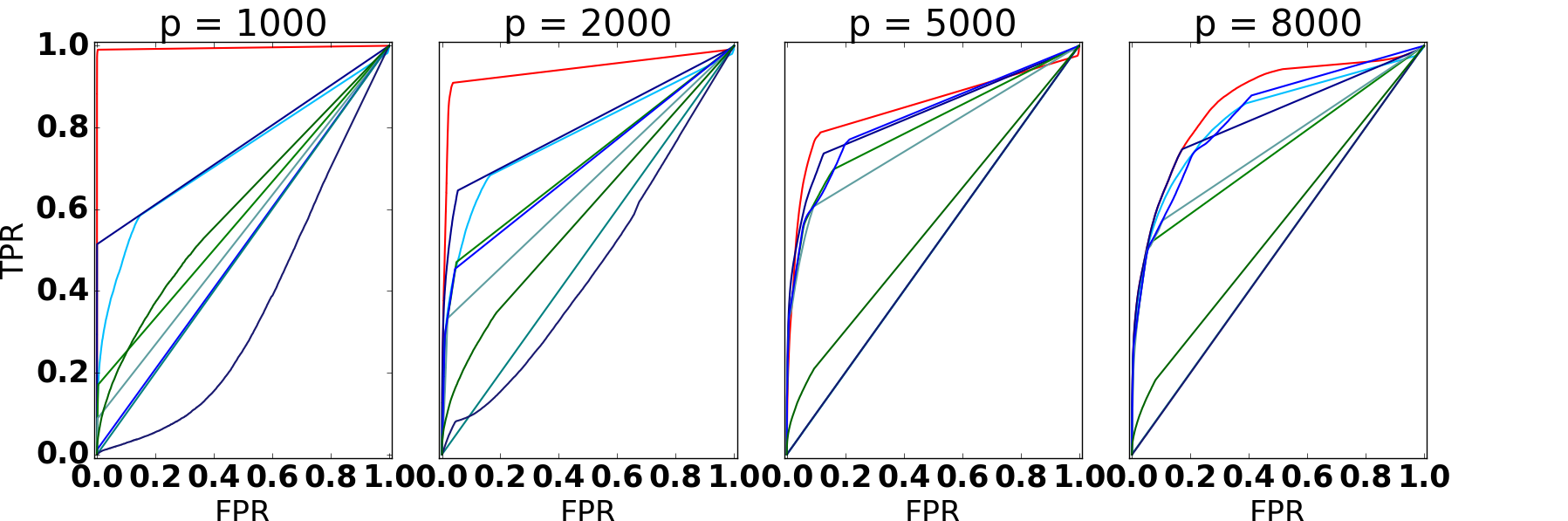}
  \label{exp}
   }\\

\subfloat[Different number of response variables]{%
  \includegraphics[width=.475\textwidth]{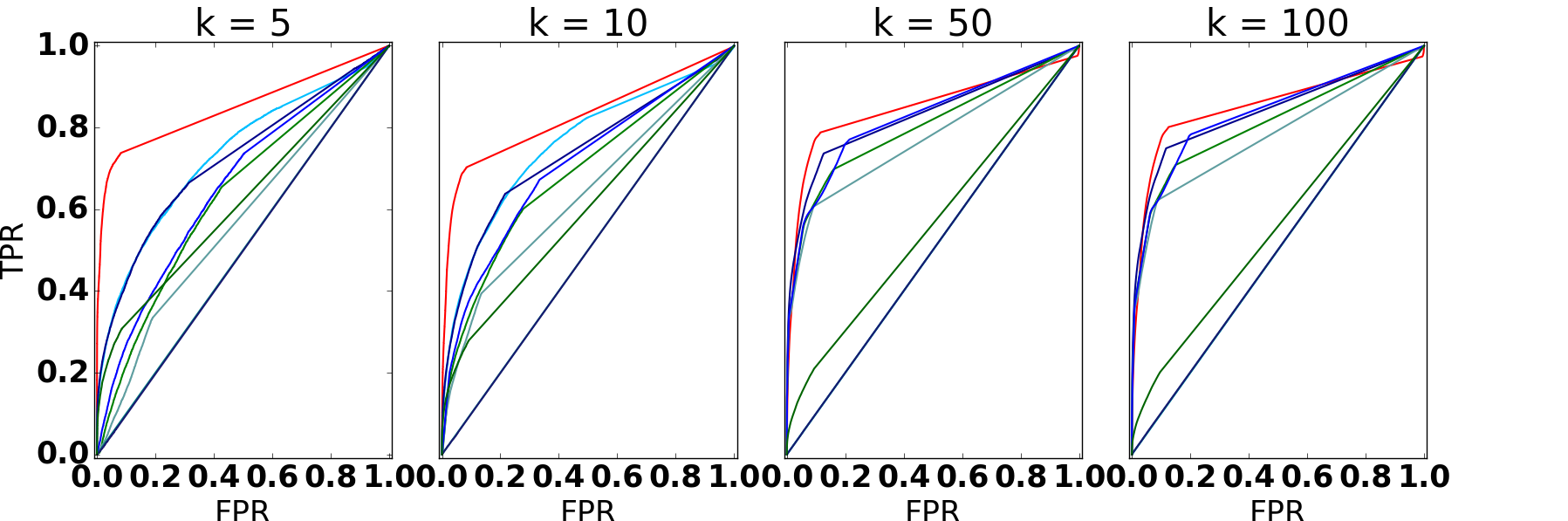}
  \label{exk}}\hfill
\subfloat[Different number of distributions]{%
  \includegraphics[width=.475\textwidth]{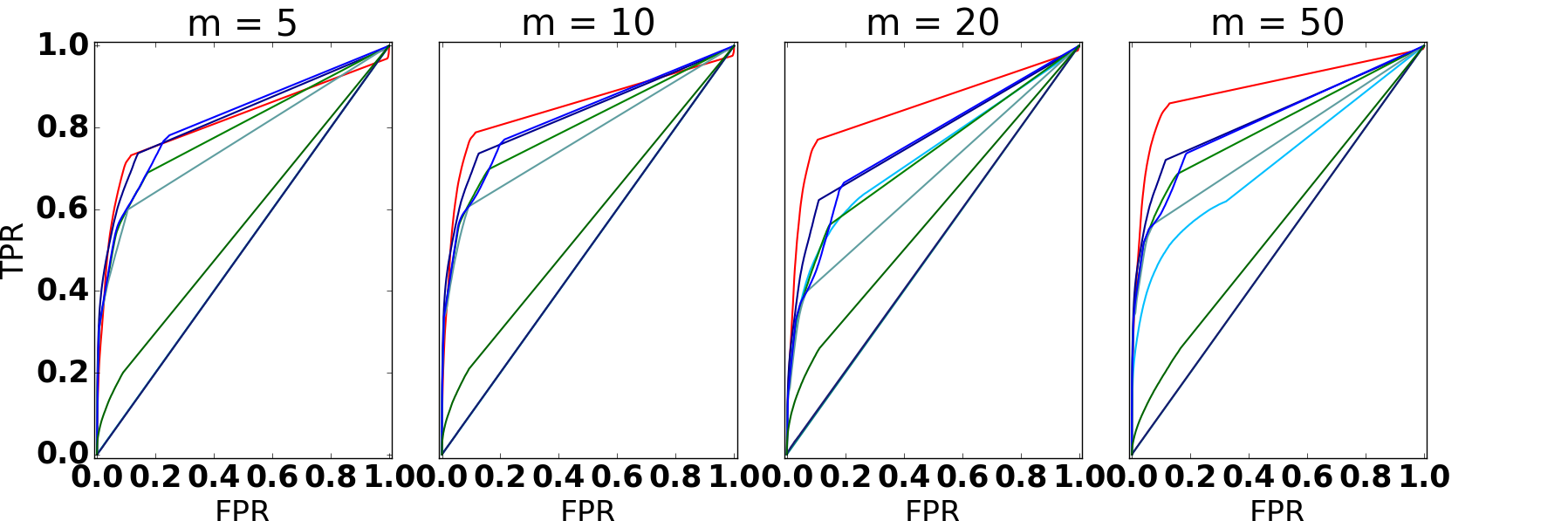}
  \label{exg}}\\
  
\subfloat[Different percentage of active variables]{%
  \includegraphics[width=.475\textwidth]{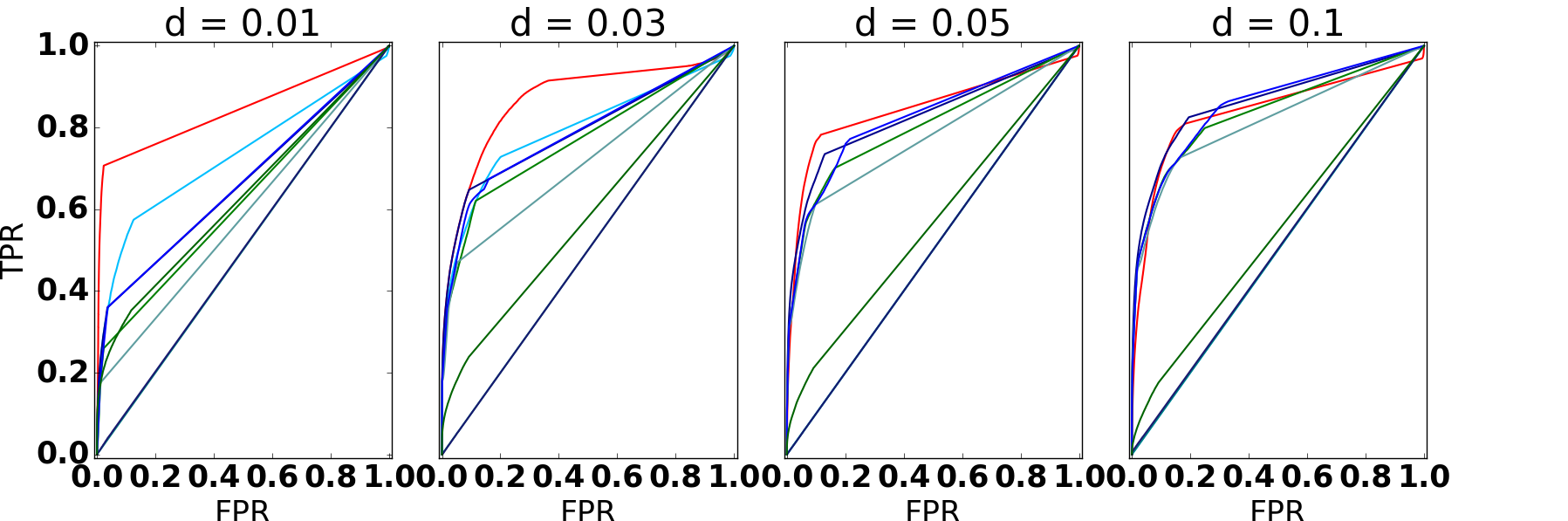}
  \label{exd}}\hfill
\subfloat[Different magnitude of variance of explanatory variables]{%
  \includegraphics[width=.475\textwidth]{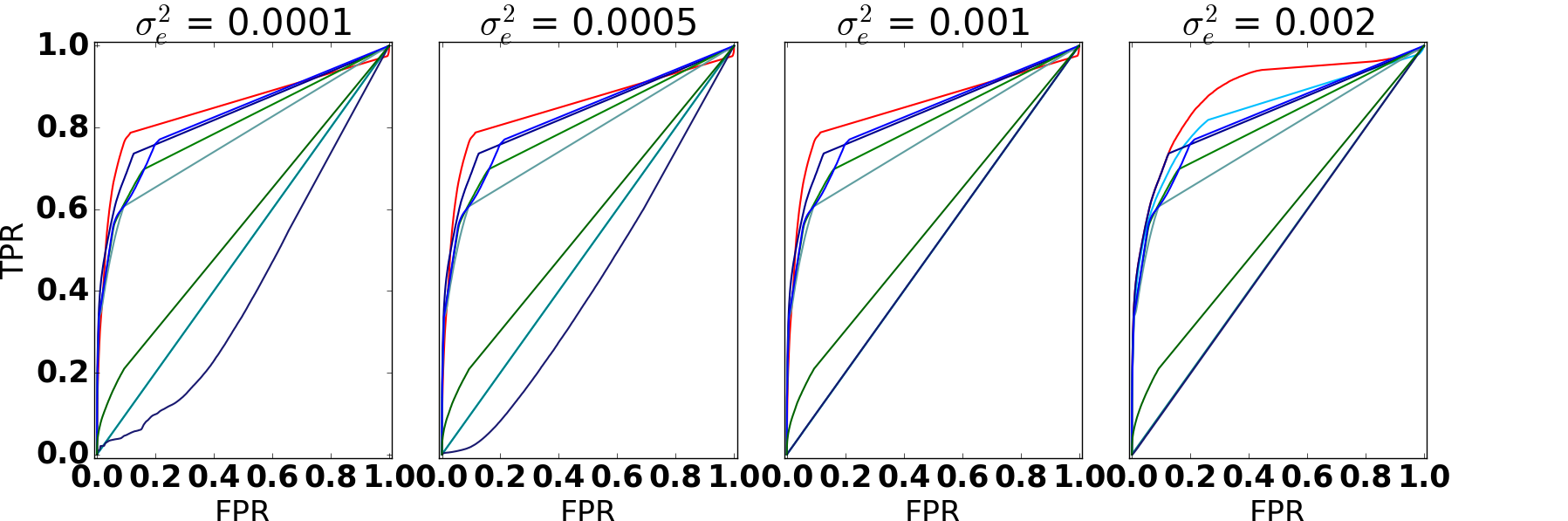}
  \label{exs}}\\
  
\subfloat[Different magnitude of variance of response variables]{%
  \includegraphics[width=.475\textwidth]{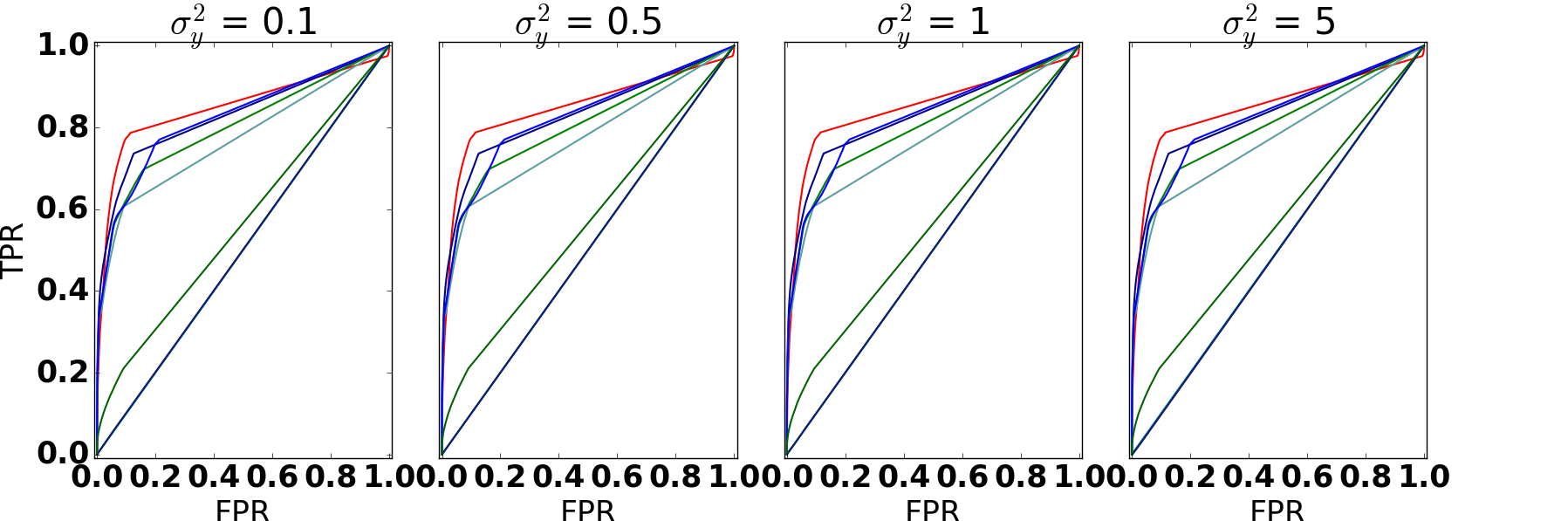}
  \label{exc}}\hfill
\subfloat[Different magnitude of noise]{%
  \includegraphics[width=.475\textwidth]{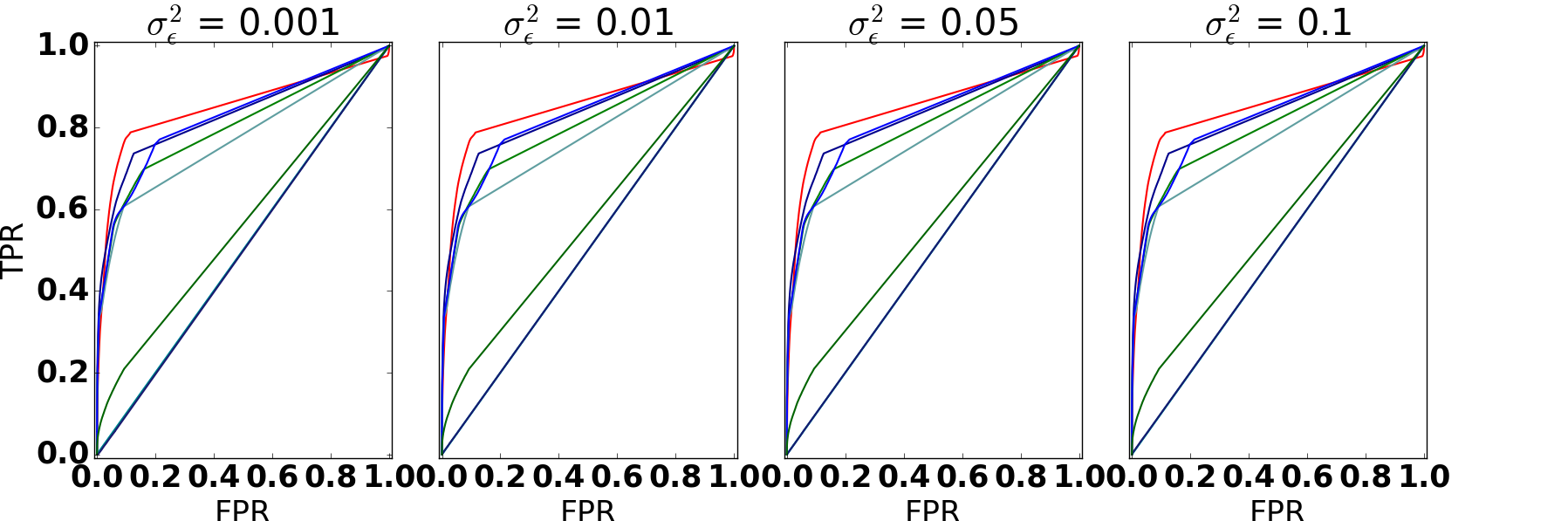}
  \label{exwe}}\\
  
\caption{ROC curves for experiments with different parameters. We show the full image of ROC curves to compare our method with previous methods. For each configuration, the reported curve is drawn over five random seeds.}
\label{fig:exp}
\end{figure*}
  
\section{SYNTHETIC EXPERIMENTS}
\label{sec:synthetic}
In this section, we evaluate the yielded results of the TgSLMM versus Tree-Lasso, LMM-Lasso and some techniques mentioned above, which is shown in the receiver operating characteristic (ROC) curves \footnote{The problem can be regarded as classification problem--identifying the active response variables from all genes. For each threshold, we select the response variables whose absolute effect sizes are greater than the threshold. If the selected explanatory variable has value above the threshold in ground truth effect size, it will be the true positive.}.
\subsection{Data Generation}
First, we simulate a sparse tree-structured vector as $\beta$. An illustrated example is shown in Figure~\ref{fig:beta_cut}. To construct $\beta$, the generation rules are listed below:

\begin{itemize}
\item The righter columns have fewer non-zero elements. 
\item The elements from righter columns have bigger value.
\item Some non-zero elements are shattered discretely in $\beta$ to increase the complexity and mimic the real situation.
\end{itemize} 

Then we generate centroids of $m$ different distributions. With $c_j$ as the centroid of $j$-th distribution, we generate explanatory variable data from a multivariate Gaussian distribution as follows: 

\begin{equation}
\label{x_j}
x_{i} \sim \mathcal{N}\,(c_j,\,\sigma_e^2{I})
\end{equation}

where $x_{i}$ denotes the $i$-th data or information bore by one individual and originates from $j$-th distribution chosen from $m$ different distributions $c$. Then we generate an immediate response vector $r$ from $X$ matrix with $\epsilon\sim\,(0 , \sigma_{\epsilon}^2)$:

\begin{equation}
\label{Re}
r=X\beta+\epsilon
\end{equation}

To get the final response matrix $Y$, we introduce a covariance matrix $K$ to simulate correlation between different responses:
\begin{equation}
\label{Y}
Y \sim \mathcal{N}\,(r,\,\sigma_y^2K)
\end{equation}
where $\sigma_y^2$ is to control the magnitude of the variance. Assuming $C$ is the matrix formed by stacking the centroids $c_j$, we choose $K=CC^T$ to simulate the correlation between observations.

Using the data generation method described earlier, our synthetic dataset can effectively mimic real-world heterogeneous datasets, capturing the desired trait relatedness.

\subsection{Experiment Results}
We assessed the ability of TgSLMM in our synthetic data sets. The experimental setting is listed in Table~\ref{tab:para}.

\begin{table}[h]\small
\caption{Default experimental setting in the simulated experiments.} 
\label{tab:para}
{\begin{tabular}
{p{1.05cm} p{0.7cm} p{5.4cm}}
\hline
Parameter & Default &Description \\\hline
$n$    &1000 &the number of data samples\\
$p$    &5000 &the number of explanatory variables\\
$k$    &50 &the number of response variables\\
$m$    &10 & the number of distributions that data originates from\\
$d$  &0.05  &the percentage of active variables\\
$\sigma_e^2$  &0.001&the magnitude of covariance of explanatory variables\\
$\sigma_y^2$ &1 &the magnitude of covariance of response variables\\
$\sigma_\epsilon^2$ &0.05 &the magnitude of noise\\
\hline
\end{tabular}}{}
\end{table} 

\begin{figure}[!b]
\centering
  \includegraphics[width=.46\textwidth]{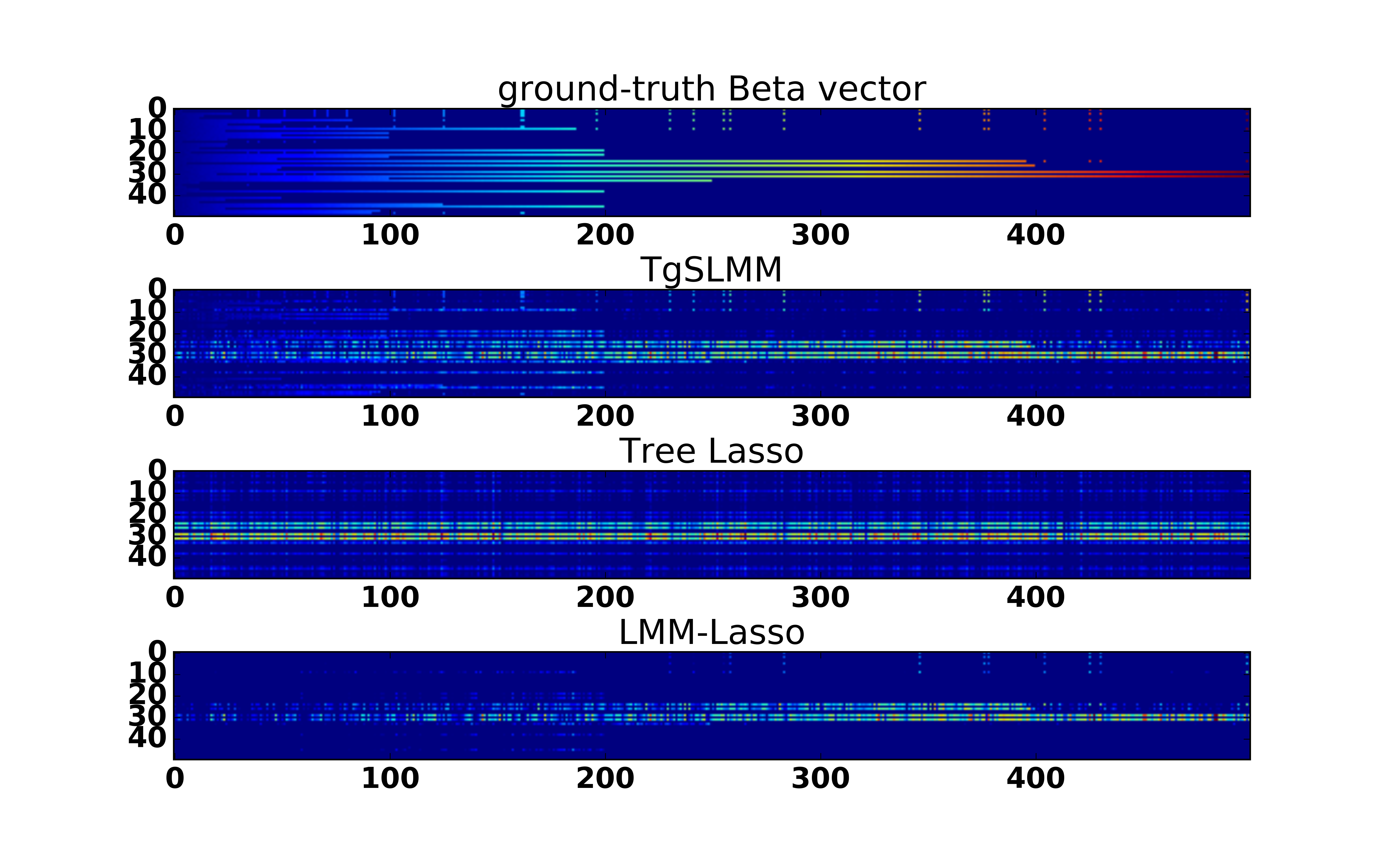}
  \caption{The yielded \boldmath{$\beta$} vectors.} 
\label{fig:beta_all}
\end{figure} 

To evaluate the performance of the proposed model in identifying active variables in different data sets, Tree-Lasso, LMM-Lasso, MCP, SCAD, Lasso\footnote{We modify Lasso, MCP, SCAD to support multidimensional data processing, the performance they yield has no observed difference. Our contributions and the details can be viewed in our codes.}, BOLT-LMM, LTMLM and LMM-Select are also tested. The baselines we choose in this paper are all highly cited and have been proven effective by many scholars. In general, our method exceeds all the other methods. The results are shown in Figure~\ref{fig:exp} considering the golden criterion ROC curves.

In Figure~\ref{fig:exp}(a) as $n$ increases, and in Figure~\ref{fig:exp}(b) as $p$ decreases, the ratio of $\frac{p}{n}$ gets smaller and the performance gets better as expected. Compared to Tree-Lasso along with other methods, our method is more robust with big data sets, which suits the real-world situation. As we increase the number of response variables in Figure~\ref{fig:exp}(c), increase the number of distributions in Figure~\ref{fig:exp}(d), or decrease the proportion of active variables in $\beta$ as Figure~\ref{fig:exp}(e), the problem becomes more challenging. Figure~\ref{fig:exp}(f) and Figure~\ref{fig:exp}(g) show that our method is more flexible to different magnitudes of covariance of explanatory variables and response variables. In Figure~\ref{fig:exp}(e), we notice that when the proportion of active variables in $\beta$ is large, the performance of TgSLMM and LMM-Lasso is similar. However, it contradicts the background of our research that the active variables should be sparse among data. Through our experiments, it is hard for Tree-Lasso to identify the active variables on high dimensional heterogeneous data.

TgSLMM also performs best in most cases in the figure of Precision-Recall curves. These figures are shown in Appendix.

\subsection{Analysis of Yielded \boldmath{$\beta$} and \boldmath{$Y$}}
\begin{figure}[!htbp]
\centering
  \includegraphics[width=.46\textwidth]{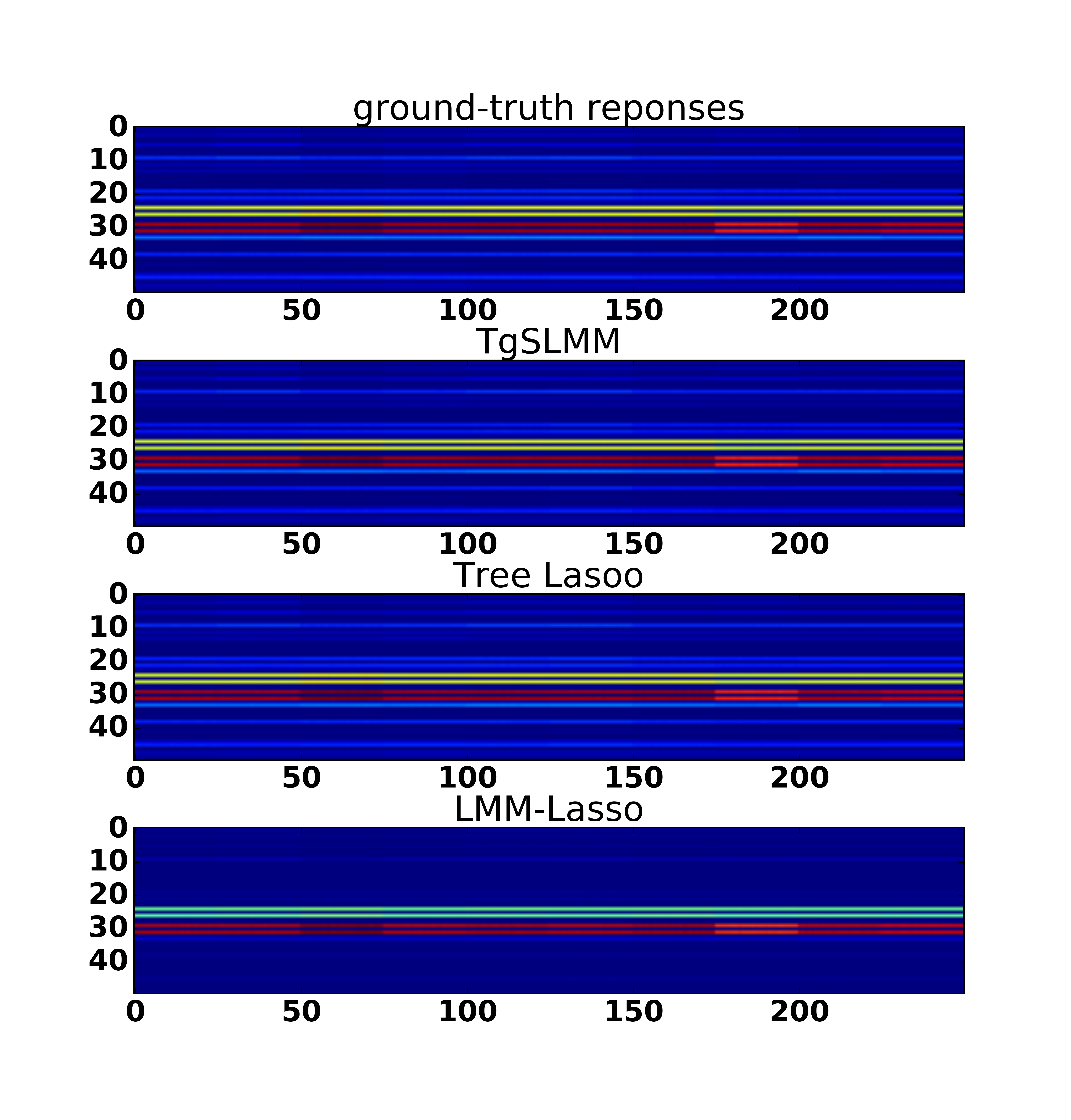}
  \caption{The simulated responses matrices and the predicted responses results by different models.}  
\label{fig:y_all}
\end{figure}

We use the same experimental setting\footnote{Other parameters are as follow: $m$ is 10; $d$ is 0.05; $\sigma_e^2$ is 0.001; $\sigma_y^2$ is 1; $\sigma_\epsilon^2$ is 0.05.} as in Figure~\ref{fig:beta_cut} to perform the ablation studies. The results are shown in Figure~\ref{fig:beta_all} and \ref{fig:y_all}.

Figure~\ref{fig:beta_all} shows that TgSLMM recovers both the values and structure of ground truth effect size, revealing the supreme ability of TgSLMM in variable selection. LMM-Lasso has trouble finding enough useful information. Trapped into the confounding factors, the Tree-Lasso discovers too many false positives. Tree-Lasso also falls short when the data set becomes complicated in the Figure~\ref{fig:exp}. Based on Figure~\ref{fig:y_all}, both prediction performance of TgSLMM and Tree-Lasso are convincing, LMM-Lasso fails as reported before. Unsurprisingly, the proposed TgSLMM also behaves the best in estimating $\beta$ with respect to mean-squared error through almost all the experimental settings. The other approaches cannot discover any meaningful information.


By using the proposed method, we are able to detect weak signals and reveal clear groupings in the patterns of associations between explanatory variables and responses and apply our method to many applications, such as variable selection, effect sizes estimation, and response prediction.

\begin{figure*}[!htbp] 
\centering
 \includegraphics[width=0.88\textwidth]{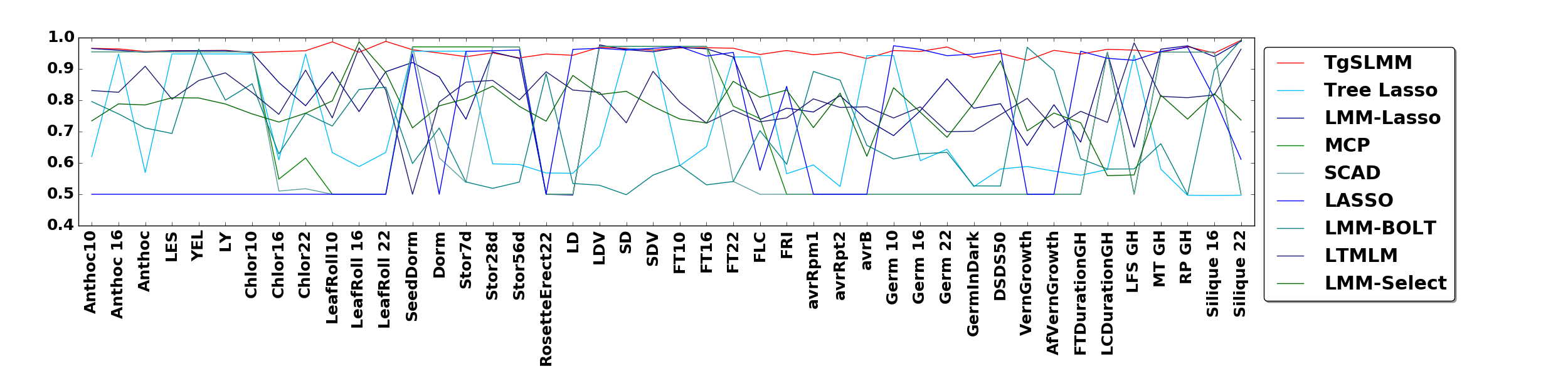}
\caption{Area under ROC curve for the 44 traits of Arabidopsis thaliana.} 
\label{fig:at_roc}
\end{figure*}

\section{REAL GENOME DATA EXPERIMENTS}
\label{sec:real}
Having shown the capacity of TgSLMM in recovering explanatory variables of synthetic data sets, we now demonstrate how TgSLMM can be used in real-world genome data and discover meaningful information. To evaluate the method, we focus on some practical data sets, Arabidopsis thaliana, Heterogeneous Stock Mice and Human Alzheimer Disease. Since Arabidopsis thaliana and Heterogeneous Stock Mice have been studied for over a decade, the scientific community has reached a general consensus regarding these species \cite{atwell2010genome}. With such authentic golden standard, we could plot the ROC curve and assess the model's performance using the area under it. However, since Alzheimer’s disease is a very active area of research with no ground truth available, we list the genetic variables identified by our proposed model and verify the top genetic variables by searching the relevant literature. 

\subsection{Data Sets}
\subsubsection{Arabidopsis Thaliana}
The Arabidopsis thaliana data set we obtained is a collection of around 200 plants, each with around 215,000 genetic variables \cite{anastasio2011source}. We study the association between these genetic variables and a set of observed responses. These plants were gathered from 27 different countries in Europe and Asia, so that geographic origin served as a potential confounding factor. For example, different sunlight conditions in different regions may affect the observed responses of these plants. We test the genetic associations between genetic variables with 44 different responses such as days to germination, days to flowering, \textit{etc}.

\subsubsection{Heterogeneous Stock Mice}
The heterogeneous stock mice data set contains measurements from around 1700 mice, with 10,000 genetic variables \cite{valdar2006genome}. These mice were raised in cages by four generations over a two-year period. In total, the mice came from 85 distinct families. The obvious confounding variable is genetic inheritance due to family relationships. We study the association between the genetic variables and a set of 27 response variables that could possibly be affected by inheritance. These 27 response variables fall into six different categories, relating to the glucose level, insulin level, immunity, EPM, FN and OFT respectively.

\subsubsection{Human Alzheimer Disease}
We use the late-onset Alzheimer's Disease data provided by Harvard Brain Tissue Resource Center and Merck Research Laboratories \cite{zhang2013integrated}. It consists of measurements from 540 patients with 500,000 genetic variables. We test the association between these genetic variables and 28 responses corresponding to a patient's disease status of Alzheimer's disease.

\subsubsection{Preprocessing of Real Genomic Data}
Each element of the explanatory variables $X$ takes values from \{0,1\} according to the number of minor alleles frequency (MAF) at the given locus in each individual. We also standardized the traits data, according to the analysis and statistics law. In the experiments, we found that the standardizing process is very crucial to the performance of the model.

\begin{figure*}[!htbp]
\centering
\includegraphics[width=0.88\textwidth]{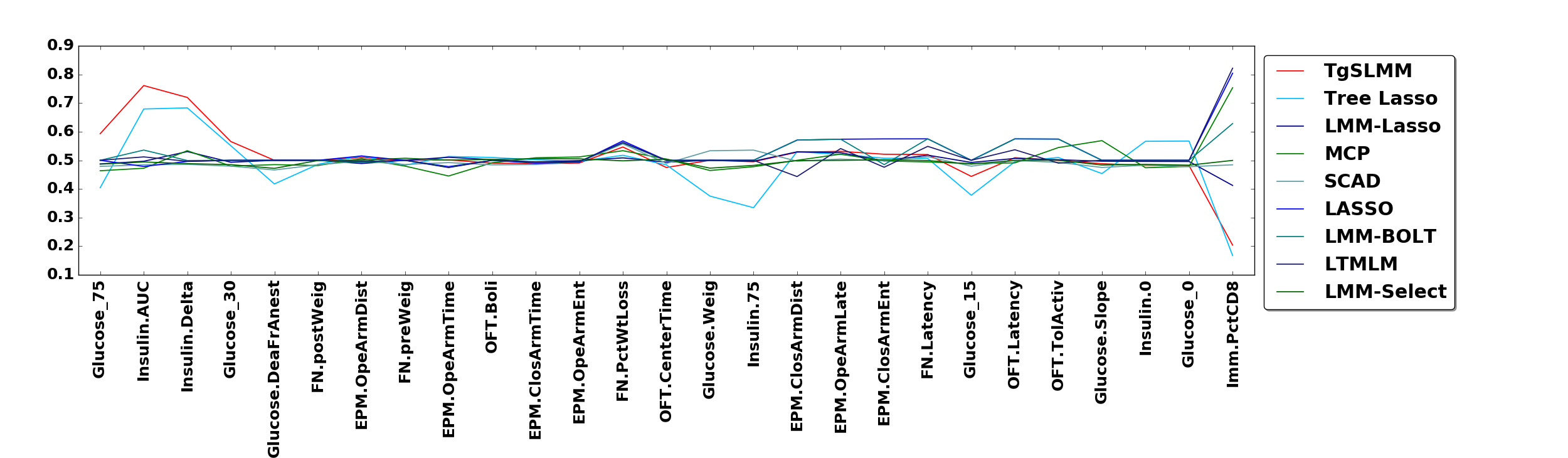}
\caption{Area under ROC curve for the 27 traits of mice.}
\label{fig:mice_roc}
\end{figure*}

\subsection{Arabidopsis Thaliana}
Since we have access to a validated gold standard of the Arabidopsis thaliana data set, we compare the alternative algorithms in terms of their ability in recovering explanatory variables with a true association. Figure~\ref{fig:at_roc} illustrates the area under the ROC curve for each response variable for Arabidopsis thaliana. By analyzing the results, we conclude that TgSLMM equals or exceeds the other methods for all of responses. TgSLMM allows for dissecting individual explanatory variable effects from global genetic effects driven by population structure.

Further, we simply apply linear regression and cross-validation to evaluate the proposed model's ability of response prediction versus all the algorithms. Using the explanatory variables our proposed method selects, 61.4\% of prediction for Arabidopsis thaliana is better than using origin data set, 56.8\% is better than using the data after employing Tree-Lasso, 79.5\% is better than applying LMM-Lasso, 84.1\% is better than MCP and SCAD, 66.0\% is better than Lasso, 91.0\% is better than LMM-BOLT, 56.7\% is better than LTMLM. Our method only works worse than LMM-Select while considering prediction.

\subsection{Heterogeneous Stock Mice}
For Heterogeneous Stock Mice data set, ground truth is also available so that we could evaluate the methods based on the area under their ROC Curve as Figure~\ref{fig:mice_roc}. TgSLMM behaves as the best one on 22.2\% of the traits and achieves the highest ROC area for the whole data set as 0.627. The second best model is MCP with the area of 0.604. The areas under ROC of Tree-Lasso, Lasso and SCAD are 0.582, 0.591 and 0.590 respectively. The areas of the remaining models are all around 0.5, showing little ability to process such complex data sets. On traits Glucose\_75, Glucose\_30, Glucose.DeadFromAnesthetic, Insulin.AUC, Insulin.Delta and FN.postWeight, our method TgSLMM behaves the best. The results are interesting: the left side of the figure mostly consists of traits regarding glucose and insulin in the mice, while the right side of the figure consists of traits related to immunity. This raises the inspiring question of whether or not immune levels in stock mice are largely independent of family origin. 

\subsection{Human Alzheimer Disease}
Finally, we proceed to the Human Alzheimer’s Disease data set and report the top 99 genetic variables our model discovered in Table~\ref{table:result} to foster further research.

\begin{table}[h]\small
\caption{Discovered Genetic Variables with TgSLMM.}
\label{table:result}
\centering
\begin{tabular}
{p{0.4cm}p{1.4cm}|p{0.4cm}p{1.4cm}|p{0.4cm}p{1.4cm}}
\hline
  Rank & SNP & Rank & SNP &Rank & SNP  \\
  \hline

1	&rs30882	&34	&rs10775247	&67	&rs3129317\\
2	&rs10027921	&35	&rs13129773	&68	&rs7588938\\
3	&rs12641981	&36	&rs9598119	&69	&rs551499\\
4	&rs12506805	&37	&rs3767444	&70	&rs11776648\\
5	&rs684240	&38	&rs10904362	&71	&rs635154\\
6	&rs16844380	&39	&rs7516808	&72	&rs10965041\\
7	&rs6431428	&40	&rs7039420	&73	&rs2301230\\
8	&rs12509328	&41	&rs11673516	&74	&rs164415\\
9	&rs7783626	&42 &rs6552578	&75	&rs11203999\\
10	&rs11848278	&43	&rs7043499	&76	&rs7302430\\
11	&rs10897029	&44	&rs11642659	&77	&rs31331\\
12	&rs464906	&45	&rs6571869	&78	&rs1421203\\
13	&rs874404	&46	&rs12648715	&79	&rs11953877\\
14	&rs4421632	&47	&rs1475668	&80	&rs11607862\\
15	&rs6086773	&48	&rs10247315	&81	&rs6762993\\
16	&rs4882754	&49	&rs467089	&82	&rs7969169\\
17	&rs2272445	&50	&rs10512516	&83	&rs12523589\\
18	&rs12410705	&51	&rs17077288	&84	&rs6544170\\
19	&rs4578488	&52	&rs6852162v	&85	&rs8107465\\
20	&rs3887171	&53	&rs2492303	&86	&rs509936\\
21	&rs2298955	&54	&rs12664420	&87	&rs659505\\
22	&rs1998933	&55	&rs993312	&88	&rs2415449\\
23	&rs17467420	&56	&rs4890939	&89	&rs17173637\\
24	&rs7629705	&57	&rs1979687	&90	&rs2519126\\
25	&rs27162	&58	&rs4702249	&91	&rs11605879\\
26	&rs4740820	&59	&rs10233816	&92	&rs812462\\
27	&rs1495805	&60	&rs6118709	&93	&rs13250449\\
28	&rs7916633	&61	&rs2798639	&94	&rs1080310\\
29	&rs13221797	&62	&rs12345602	&95	&rs4570478\\
30	&rs429536	&63	&rs4766333	&96	&rs4787760\\
31	&rs7668750	&64	&rs3814391	&97	&rs2682585\\
32	&rs1463118	&65	&rs1008411	&98	&rs2854439\\
33	&rs1163825	&66	&rs12815078	&99	&rs9301747\\
\hline
\end{tabular}
\end{table}

Due to space limitation, we only verify the top 10 reported genetic variables with prior research. The $1^{st}$ discovered genetic variable is corresponded to \textit{apoB} gene, which can influence serum concentration \cite{caramelli1999increased} in Alzheimer’s disease \cite{williams2020lipid}. The $2^{nd}$ one is associated with \textit{ARHGAP10} gene (also called \textit{GRAF2}), which is an important paralogon of \textit{ARHGAP26} that closely related to the Alzheimer's disease \cite{wang2022associations} and affects the developmentally regulated expression of the \textit{GRAF} proteins that promote lipid droplet clustering and growth, and is enriched at lipid droplet junctions \cite{hasler2014graf1a,williams2020lipid}. The $3^{rd}$ SNP \textit{GNPDA2} is discovered to show the environment and gene association with obesity. They have impact on neurodegenerative and neurodevelopmental diseases \cite{flores2020environment}. The $4^{th}$ SNP is expressed by the \textit{SYNPO2}, which influences hypercholesterolemia or hypertension that has a identified a link between cognitive deficits \cite{loke2017global}. The $6^{th}$ SNP known as the \textit{LY75}, has close relation with the significantly differentially expression in the time-series paired analysis involving \textit{APOE4} carriers and non-carriers, which could affect Alzheimer’s disease \cite{luckett2023longitudinal}. The $7^{th}$ genetic variable is associated with \textit{AGAP1}. \textit{AGAP1} can regulate membrane trafficking, actin remodeling \cite{liu2006novel} and is reported to be associated with Alzheimer's disease. The $8^{th}$ one is coded by gene \textit{FAM114A1}. Biologists have found that \textit{FAM114A1} is highly expressed in the developing neocortex \cite{zhang2014amyloid}. Also, from ``the amyloid hypothesis", beta-amyloid accumulation is mainly cause Alzheimer's disease \cite{murphy2010alzheimer}. The $9^{th}$ is corresponded with gene \textit{CNTNAP2} and the direct downregulation of \textit{CNTNAP2} by \textit{STOX1A} is associated with Alzheimer's disease \cite{van2012direct}.

\section{Discussions}

\subsection{Complexity}

Since a tree associated with $L$ responses can have at most $2L-1$ nodes, it is computationally efficient and spatially economical to run TgSLMM. The complexity of TgSLMM is dependent on two parts. First, the decomposition of the random effect matrix $K$ to rotate the explanatory variable and response data is cubic cost, which determines the complexity of the first step. If we reduce the covariance $K$ to a low-rank representation calculated from a small subset of $n_s$ explanatory variables. The runtime is reduced from $O(nk^2)$ to $O(n_s^2k)$. Second, we employ a smoothing proximal gradient method that is originally developed for structured-sparsity-inducing penalties. By using the efficient method, the convergence rate of the algorithm is $O(\frac{1}{\epsilon})$, given the desired accuracy $\epsilon$ and the time complexity per iteration of the smoothing proximal gradient for the Tree-Lasso is $O(p^2k+p\sum{v\in V}|G_v|)$. Thus the overall complexity for our method is $O(n_s^2k+\frac{1}{\epsilon}\times(p^2k+p\sum{v\in V}|G_v|))$.  
  
\subsection{Runtime}
To evaluate its effectiveness and practicability, we have empirically measured the runtime on the Arabidopsis thaliana dataset mentioned in our paper. On a four-core computer (3GHz 12MB L2-Cache, 8GB Memory), TgSLMM required about 4 hours CPU time. In this paper, we show that our method is scalable to large genetic dataset.

\section{CONCLUSION}
\label{sec:conclusion}
In this paper, we aim to solve the challenging task of sparse variable selection when the data are not i.i.d. This type of situation often occurs in genomics since different batches of medical data are collected from different sources for different purposes. Due to such confounding factors, nai\"vely applying the traditional variable selection methods will result in a huge number of false discoveries. In addition to that, existing algorithms ignore the convoluted interdependency among responses, hence a joint analysis that can utilize such relatedness information in a heterogeneous data set is crucial.
To address these problems, we propose the tree-guided sparse linear mixed model for sparse variable selection. Apart from extending the recent solutions of LMM that can correct confounding factors, we can perform variable selection simultaneously further to account the relatedness between different responses. By conducting extensive experiments, we compare our method with state-of-art methods and deeply analyze how confounding factors from the high dimensional heterogeneous data set influence the capability of the model to identify active variables. We show that traditional methods easily fall into the trap of utilizing false information, whereas our proposed model outperforms other existing methods in both the synthetic data set and real genome data set. We make our source code available\footnote{https://github.com/lebronlambert/TgSLMM}.

\appendices





\section{\break SYNTHETIC EXPERIMENT RESULTS}

\subsection{The Precision-recall Curve of Synthetic Experiment}

\begin{figure*}[t!]\small
\centering
\includegraphics[width=.4\textwidth]{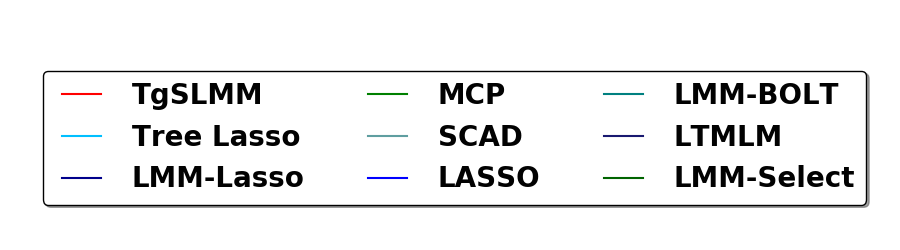}
\\

\subfloat[Different number of samples]{%
  \includegraphics[width=.48\textwidth]{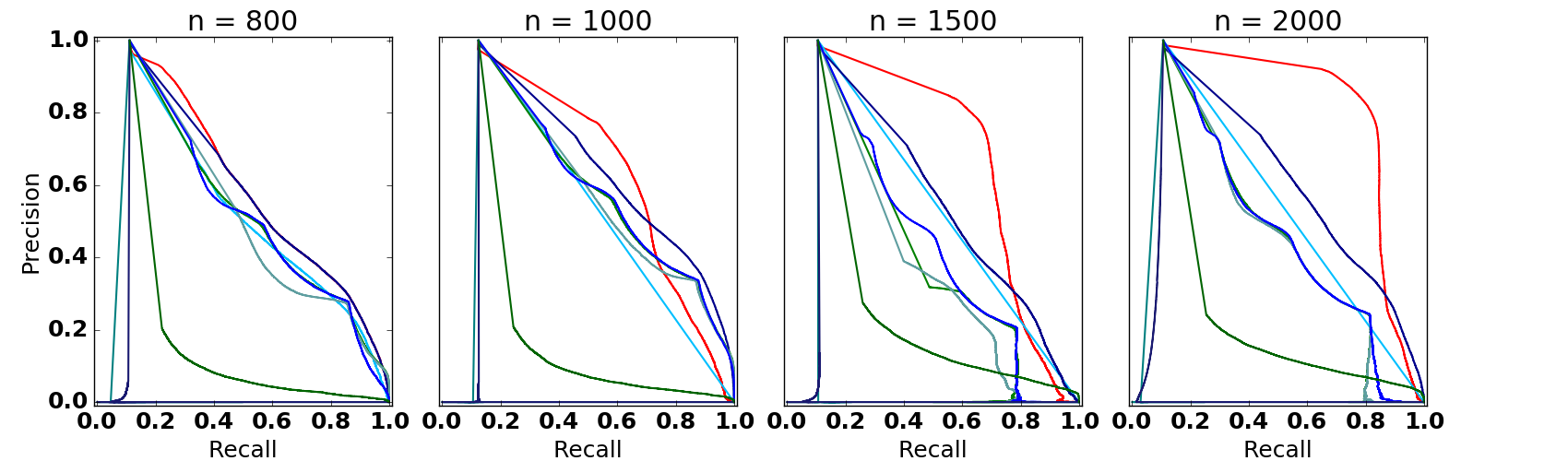}
    \label{exn_rp}}\hfill
\subfloat[Different number of explanatory variables]{%
  \includegraphics[width=.48\textwidth]{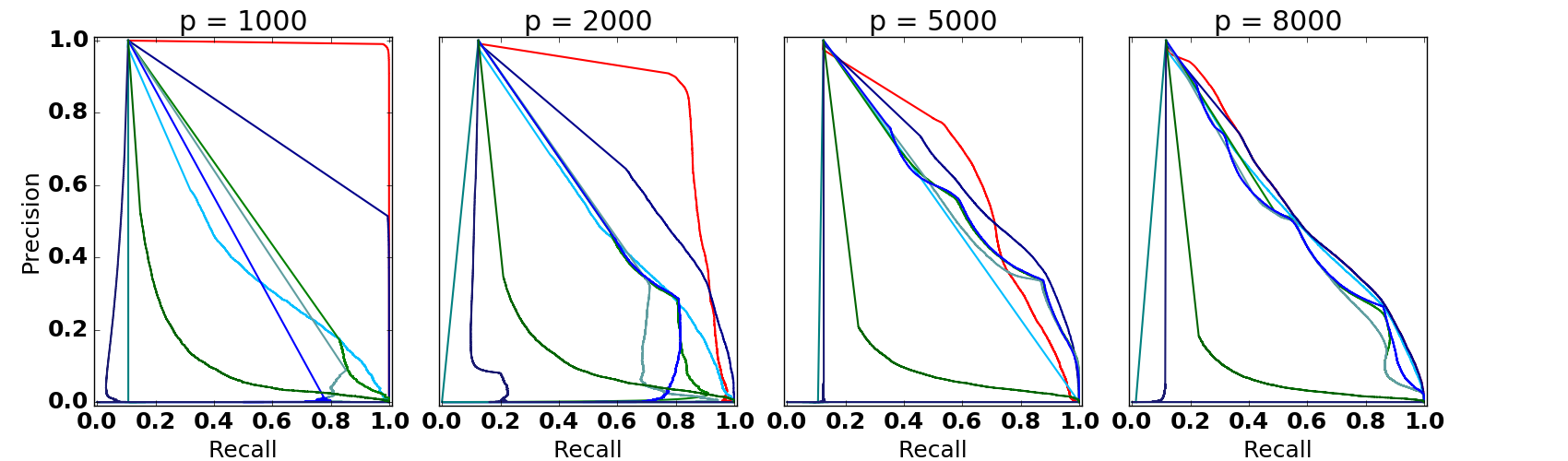}
  \label{exp_rp}
   }\\

\subfloat[Different number of response variables]{%
  \includegraphics[width=.48\textwidth]{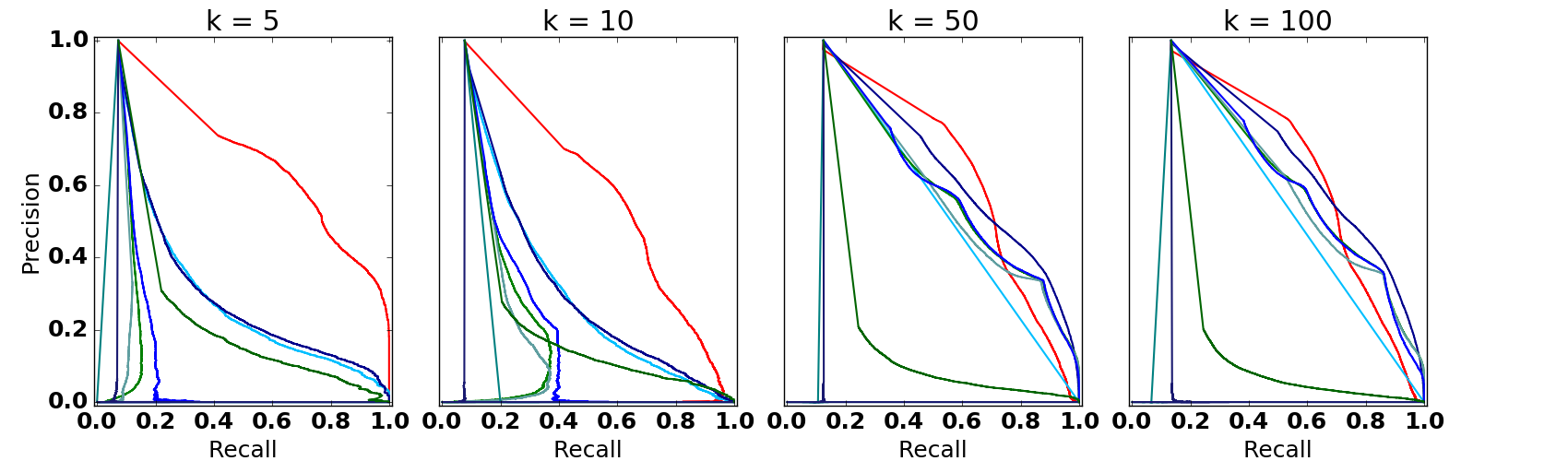}
  \label{exk_rp}}\hfill
\subfloat[Different number of distributions]{%
  \includegraphics[width=.48\textwidth]{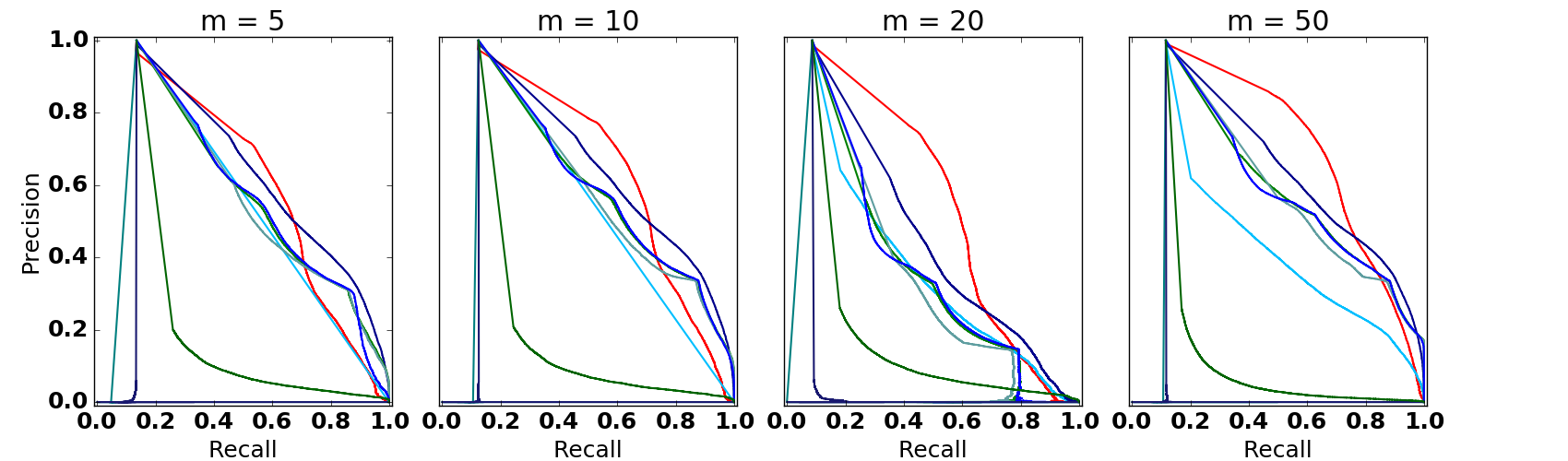}
  \label{exg_rp}}\\
  
\subfloat[Different percentage of active variables]{%
  \includegraphics[width=.48\textwidth]{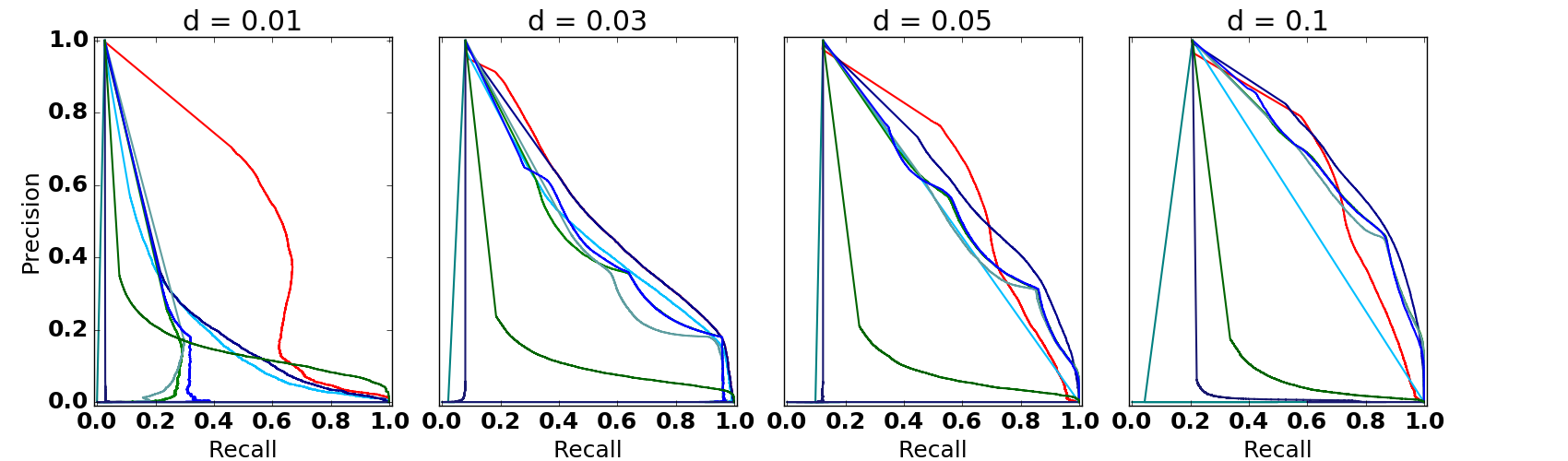}
  \label{exd_rp}}\hfill
\subfloat[Different magnitude of variance of explanatory variables]{%
  \includegraphics[width=.48\textwidth]{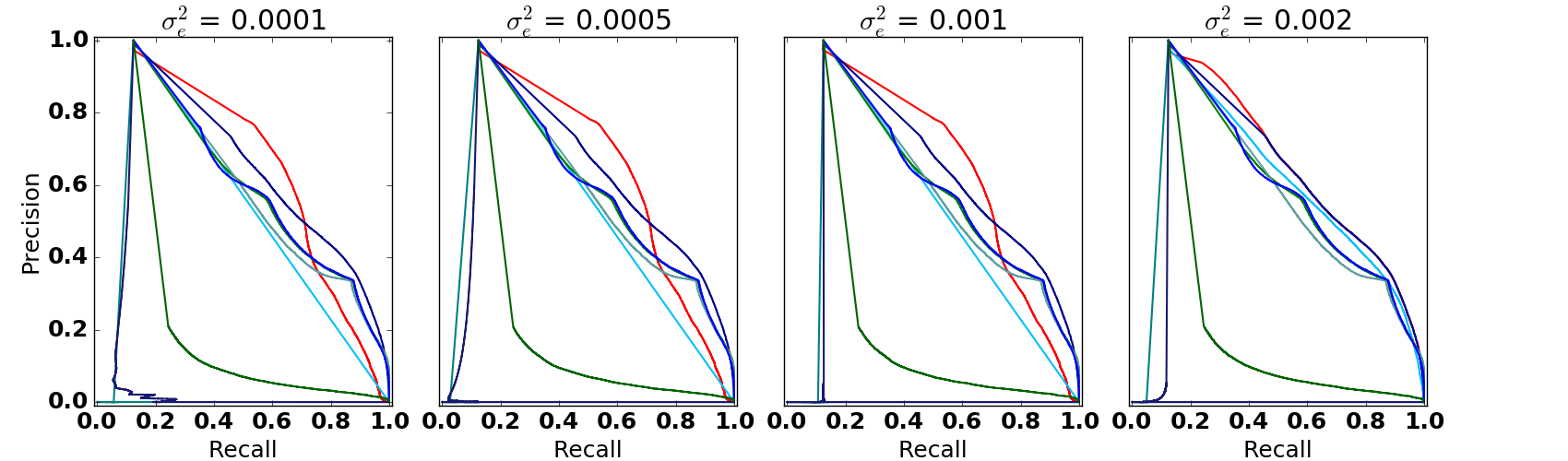}
  \label{exs_rp}}\\
  
\subfloat[Different magnitude of variance of response variables]{%
  \includegraphics[width=.48\textwidth]{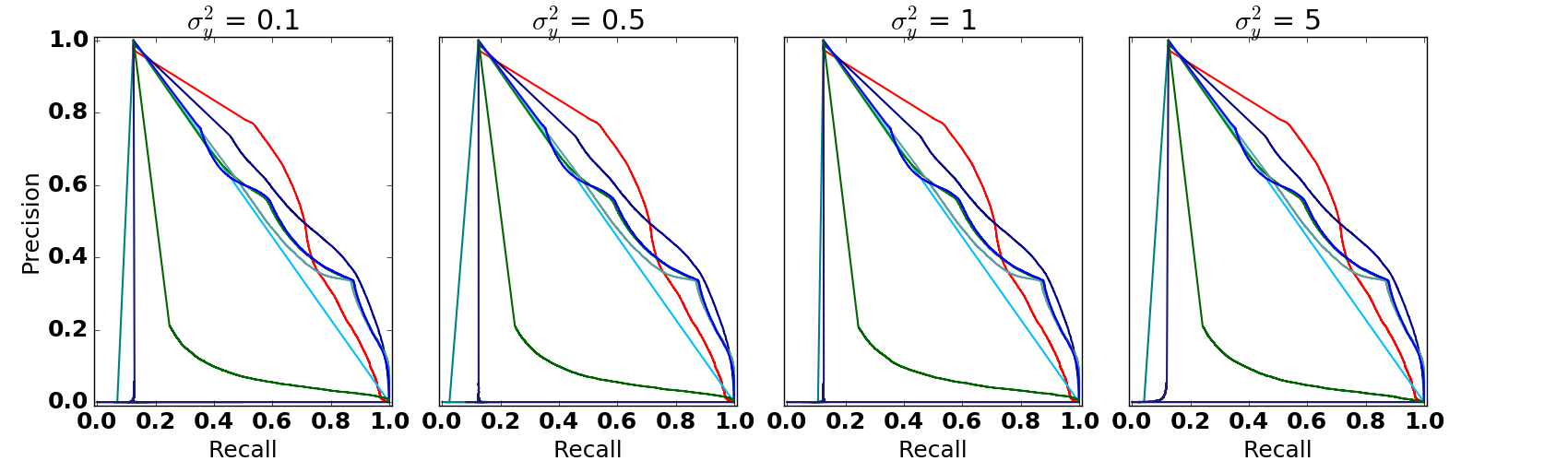}
  \label{exc_rp}}\hfill
\subfloat[Different magnitude of noise]{%
  \includegraphics[width=.48\textwidth]{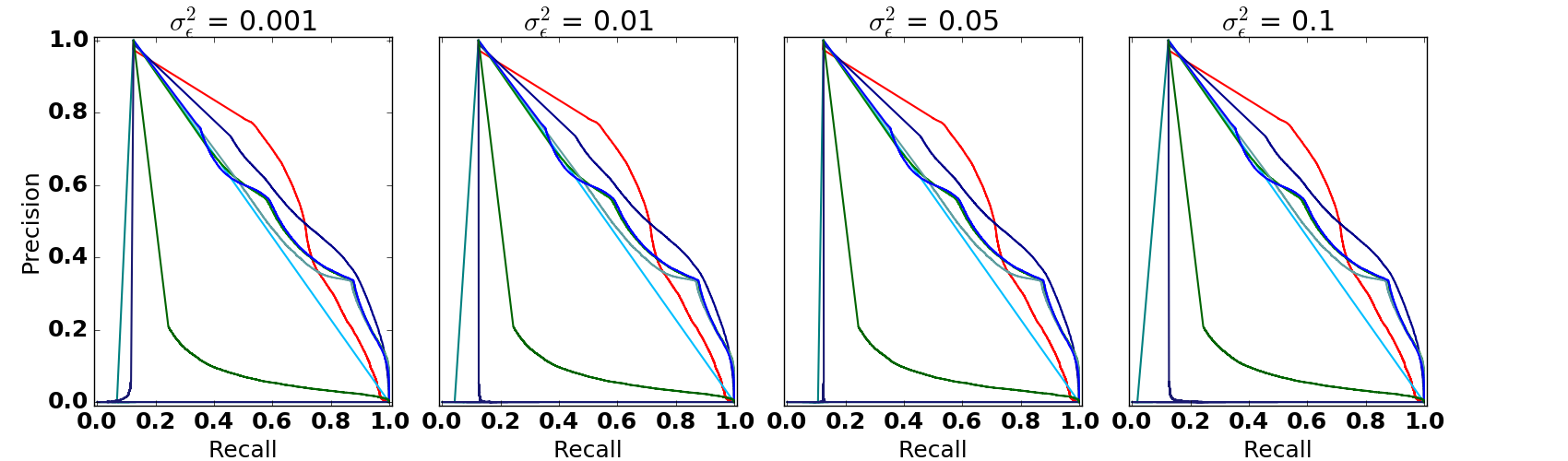}
  \label{exwe_rp}}\\
  
\caption{Precision-Recall curves for experiments with different parameters.}
\label{fig:rp}
\end{figure*}

The Figure~\ref{fig:rp} shows the full images of Precision-Recall curves in synthetic experiments to compare our method with other existing methods by using the same parameters in our paper. For each configuration, the reported curve is drawn over five random seeds. And we can see that TgSLMM behaves almost always best.

\subsection{Estimation of \boldmath{$\beta$}}
\begin{figure*}[t!]]
\includegraphics[width=.45\textwidth]{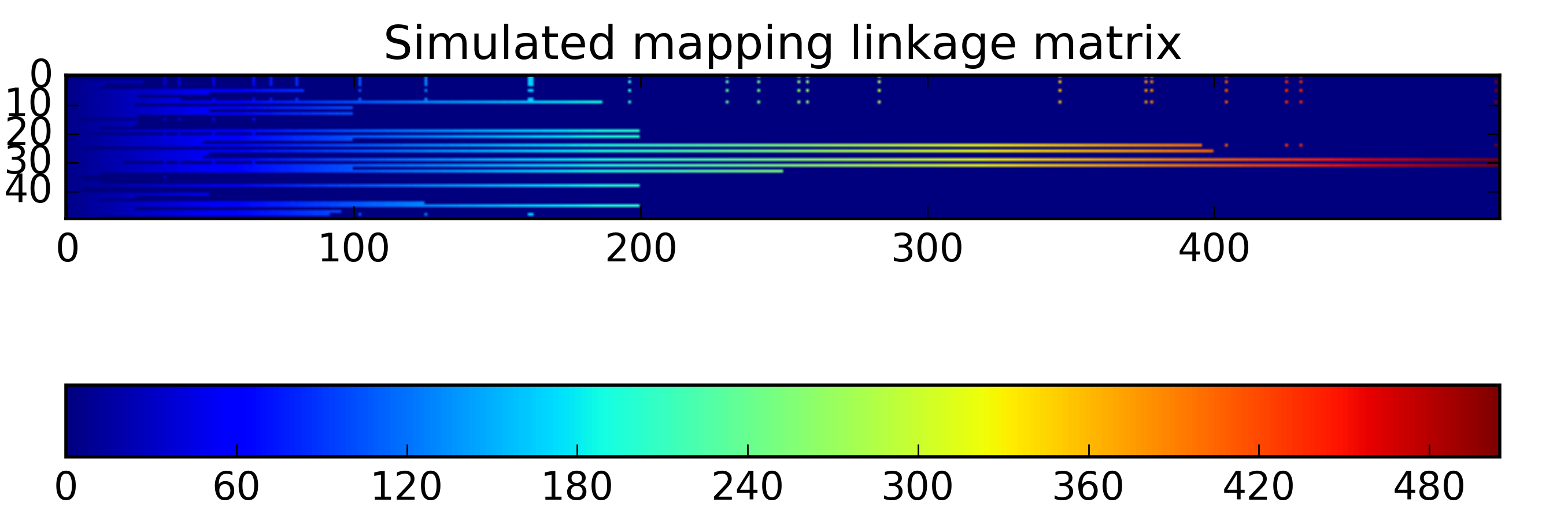}\hfill
\includegraphics[width=.45\textwidth]{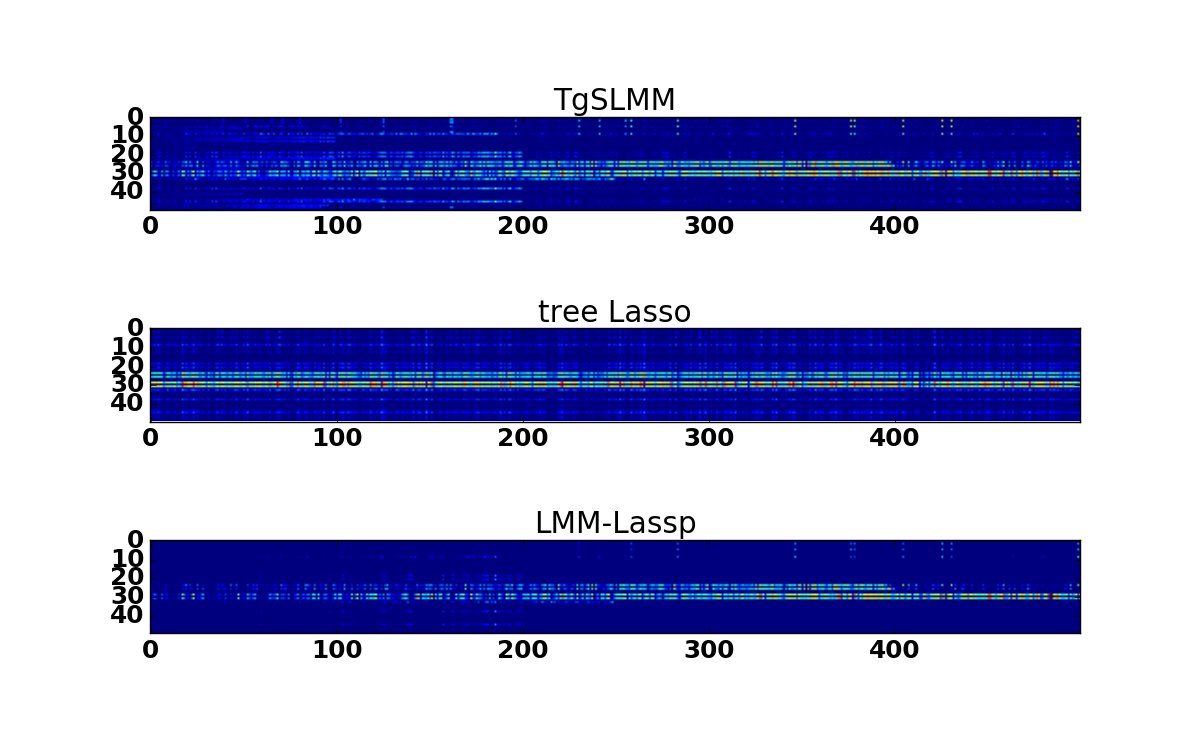}\\
\includegraphics[width=.45\textwidth]{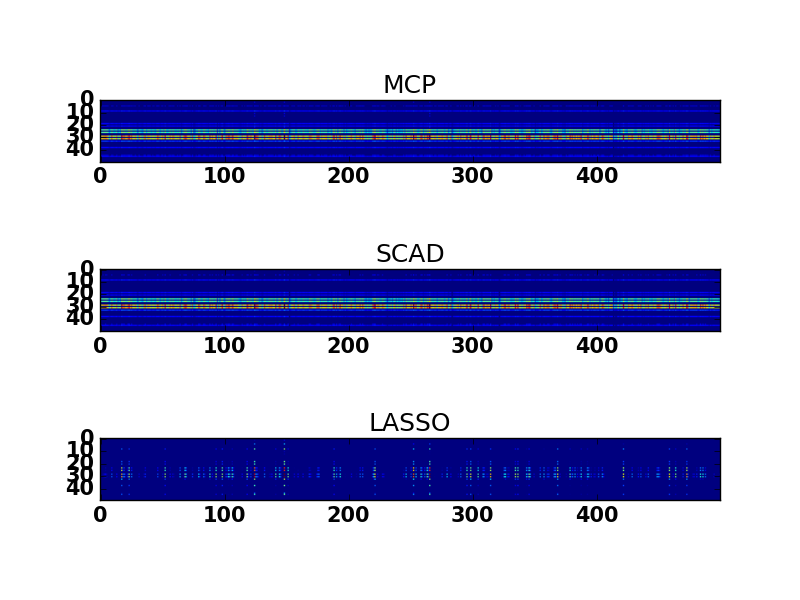}\hfill
\includegraphics[width=.45\textwidth]{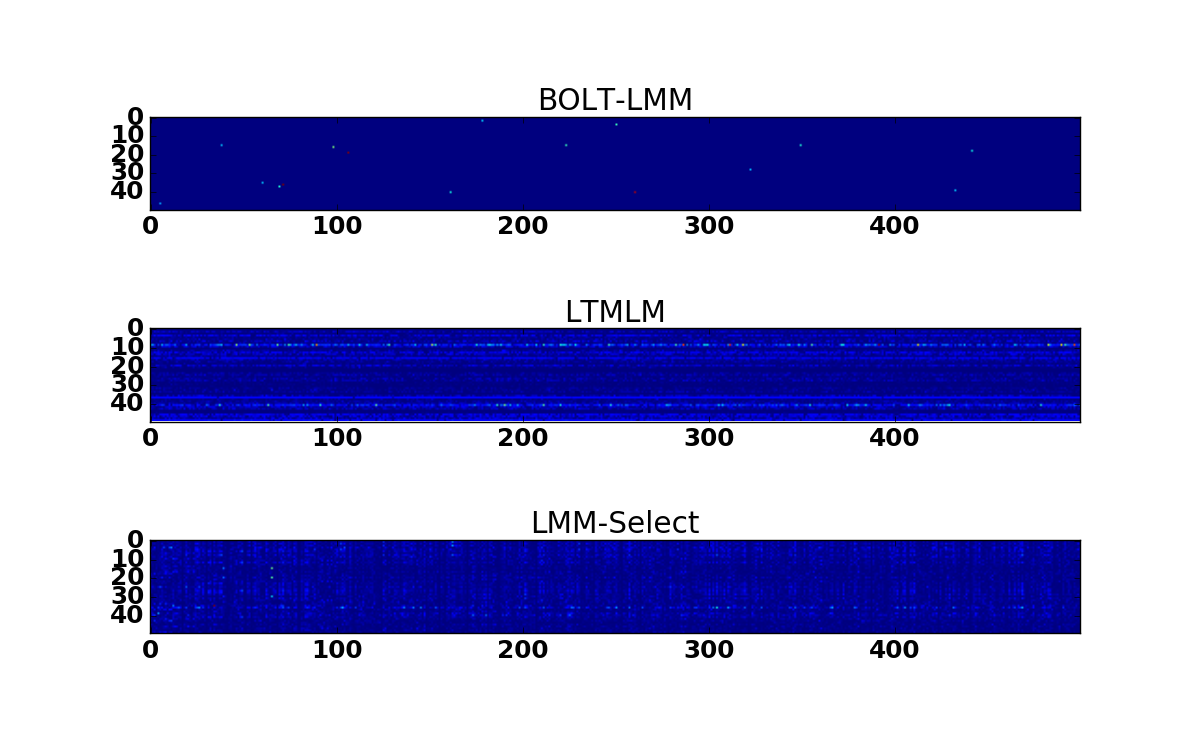}\\
\caption{The yield \boldmath{$\beta$} vector.}
\label{fig:beta}
\end{figure*}

The Figure~\ref{fig:beta} shows the $\beta$ vectors yielded by methods we used in our paper together with the ground truth $\beta$ vector generated in the synthetic experiments. The figures show that TgSLMM yields the best result with the number about 0.95 of the area under ROC curves. The area of Tree-Lasso is about 0.84, that of LMM-Lasso is around 0.71. The area under ROC of MCP, SCAD, Lasso, LMM-BOLT, LTMLM and LMM-Select is 0.81, 0.81, 0.80, 0.57, 0.50 and 0.41 respectively. 
 
\subsection{Prediction of \boldmath{$Y$}}
\begin{figure*}[t!]
\includegraphics[width=.45\textwidth]{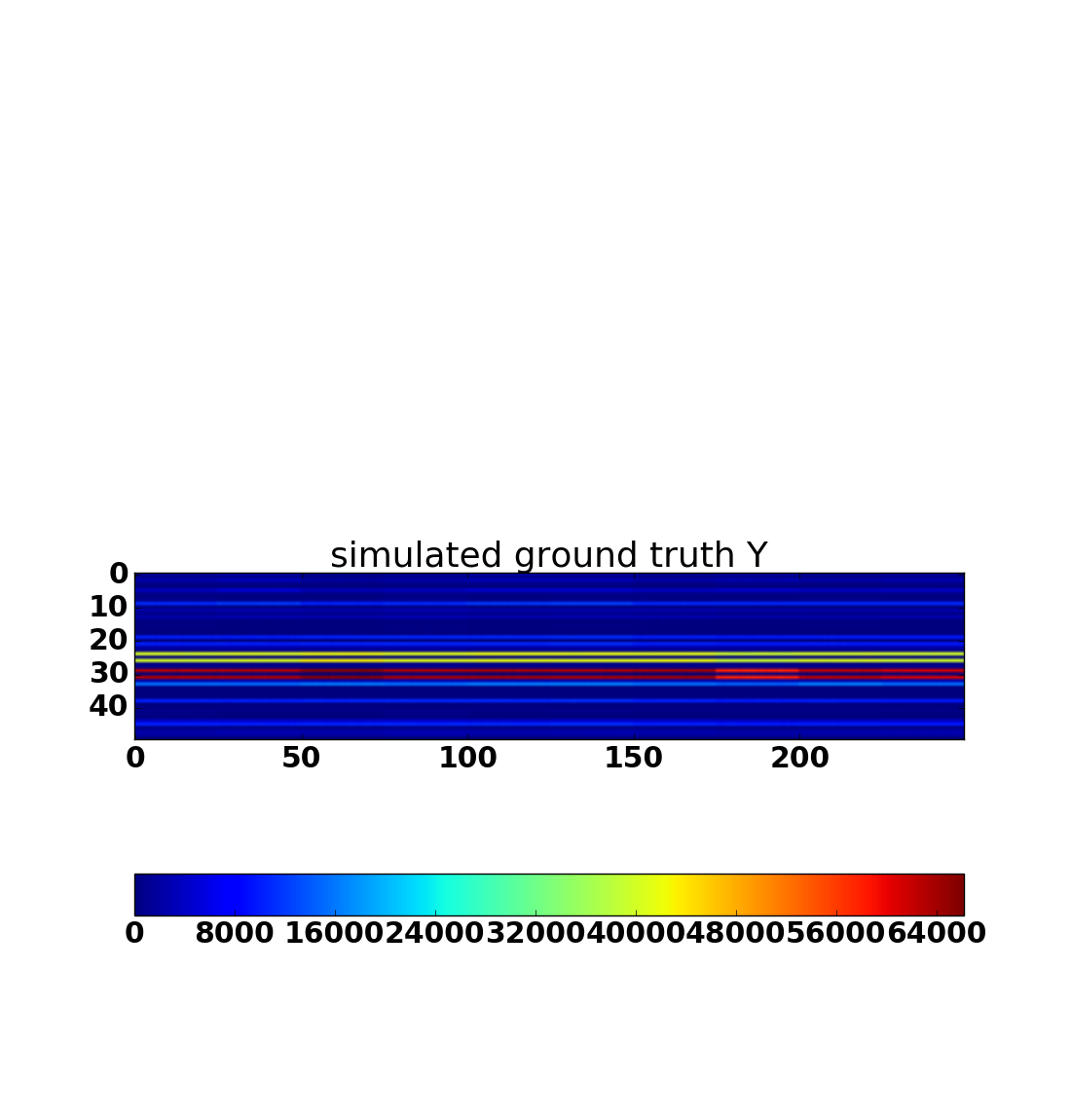}\hfill
\includegraphics[width=.45\textwidth]{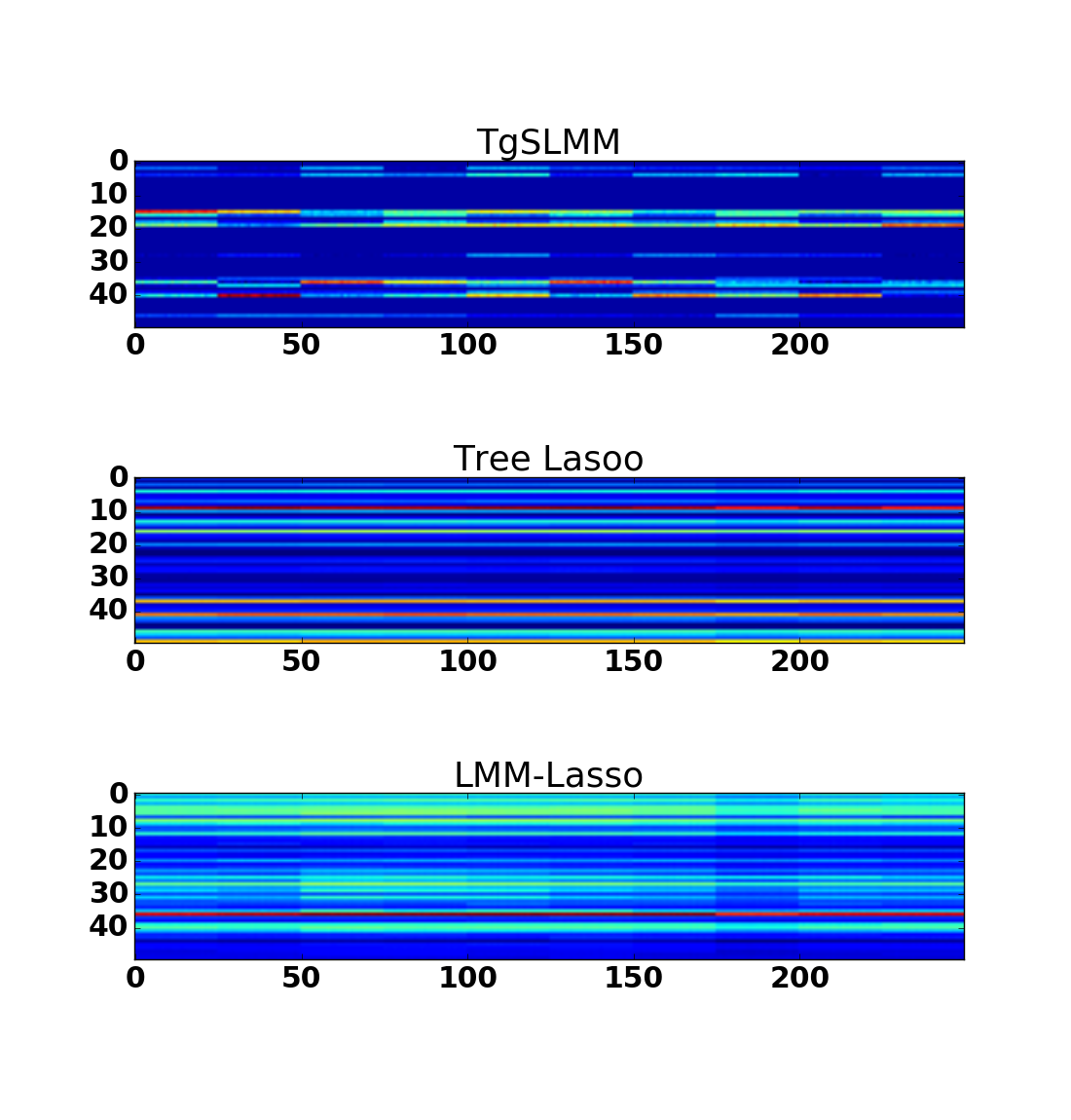}\\
\includegraphics[width=.45\textwidth]{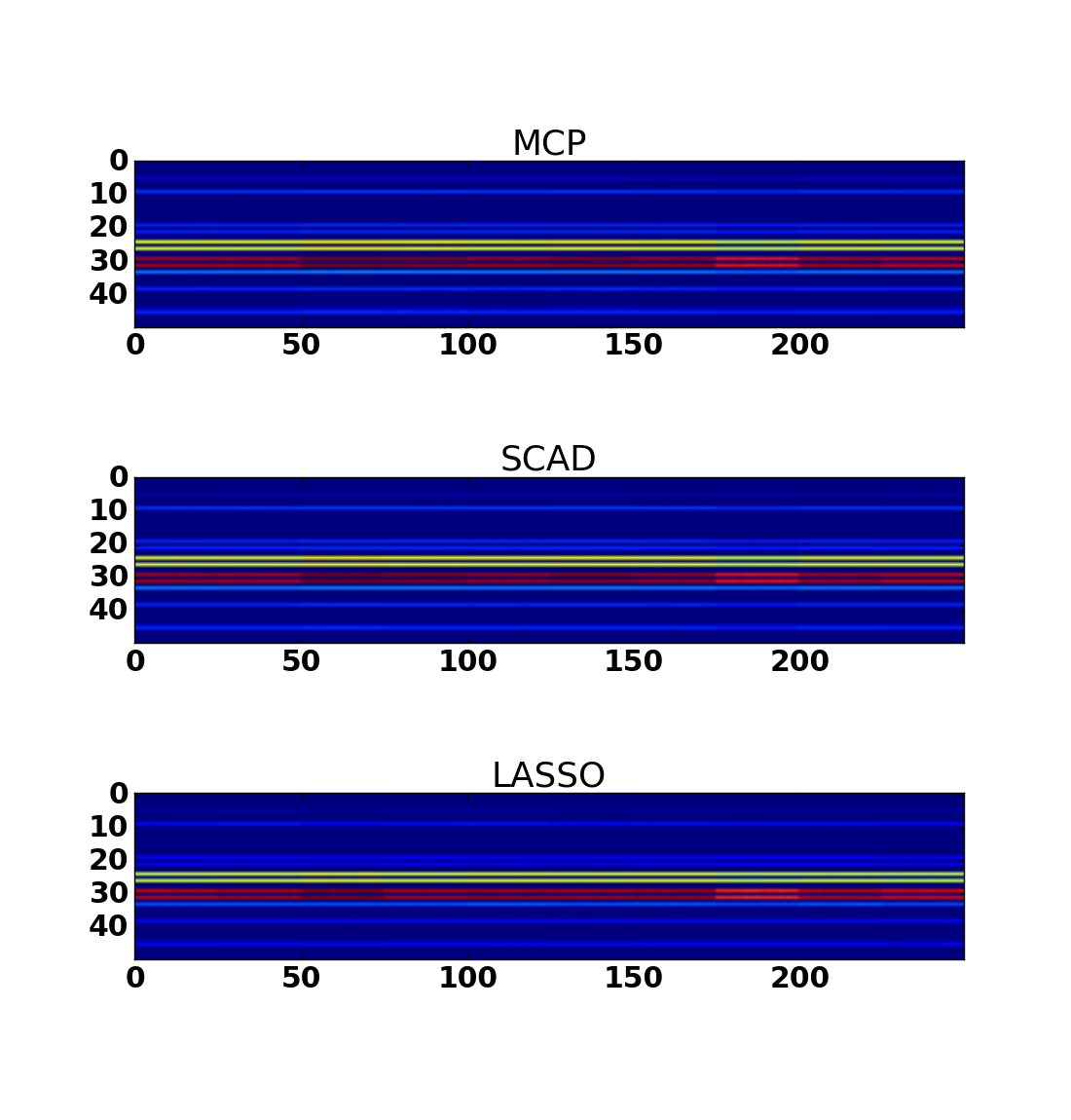}\hfill
\includegraphics[width=.45\textwidth]{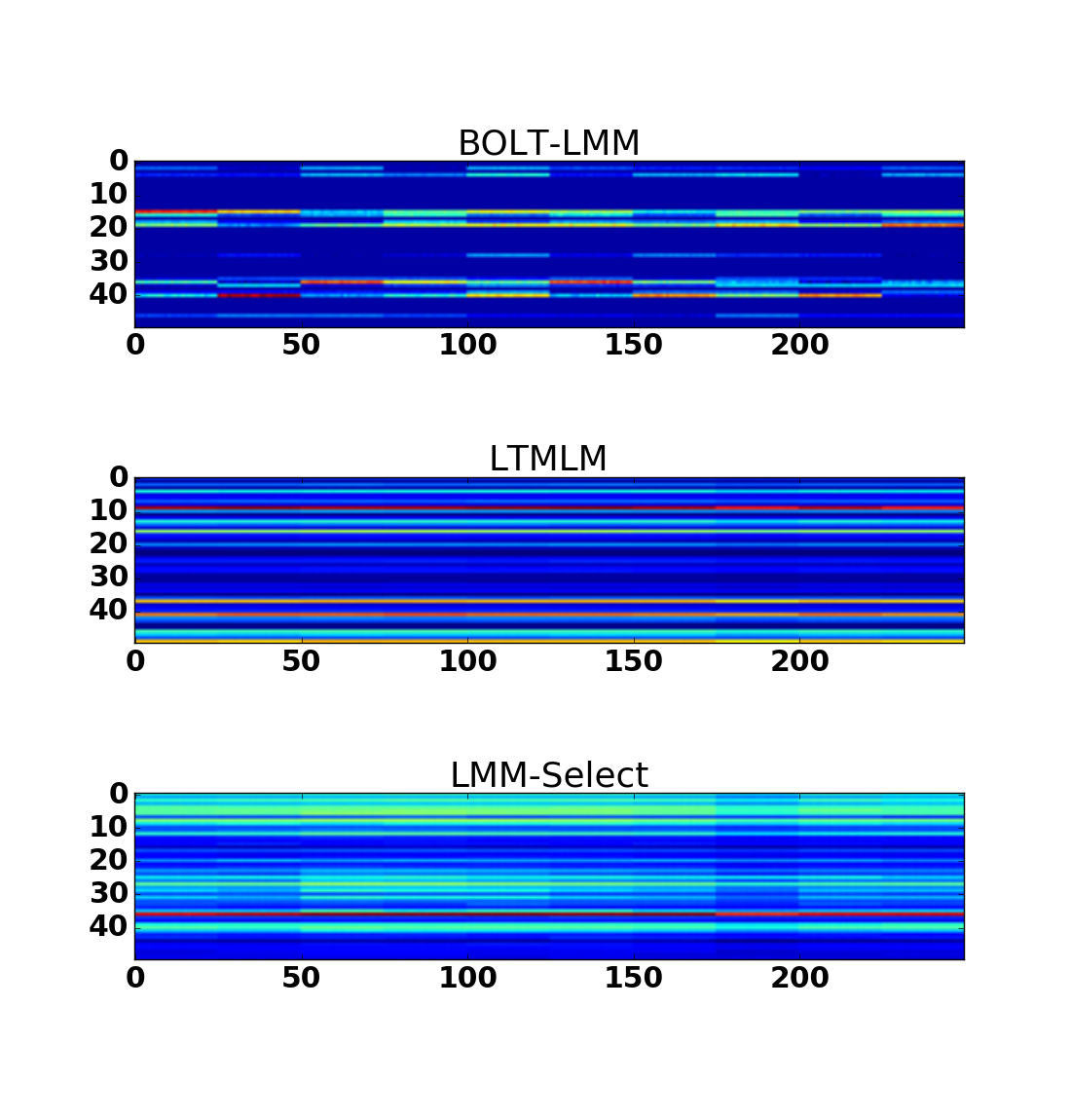}\\
\caption{The simulated responses matrix and the yield
responses results.}
\label{fig:Y_all2}
\end{figure*}

Figure~\ref{fig:Y_all2} shows the $Y$ results recovered. TgSLMM also yields the best result. \footnote{The parameters that Figure~\ref{fig:beta} and  Figure~\ref{fig:Y_all2} used are just the same experimental setting in Section 4.3 in our paper. $n$ is 250; $p$ is 500; $k$ is 50; $m$ is 10; $d$ is 0.05; $\sigma_e^2$  is 0.001; $\sigma_y^2$ is 1; $\sigma_\epsilon^2$  is 0.05; random seed is 0.} 

\subsection{The ROC Curve of Synthetic Experiment}
\begin{figure*}[t!]\small
\centering
\includegraphics[width=.4\textwidth]{PR/label_rp.png}
\\

\subfloat[Different magnitude of variance of response variables]{%
  \includegraphics[width=.48\textwidth]{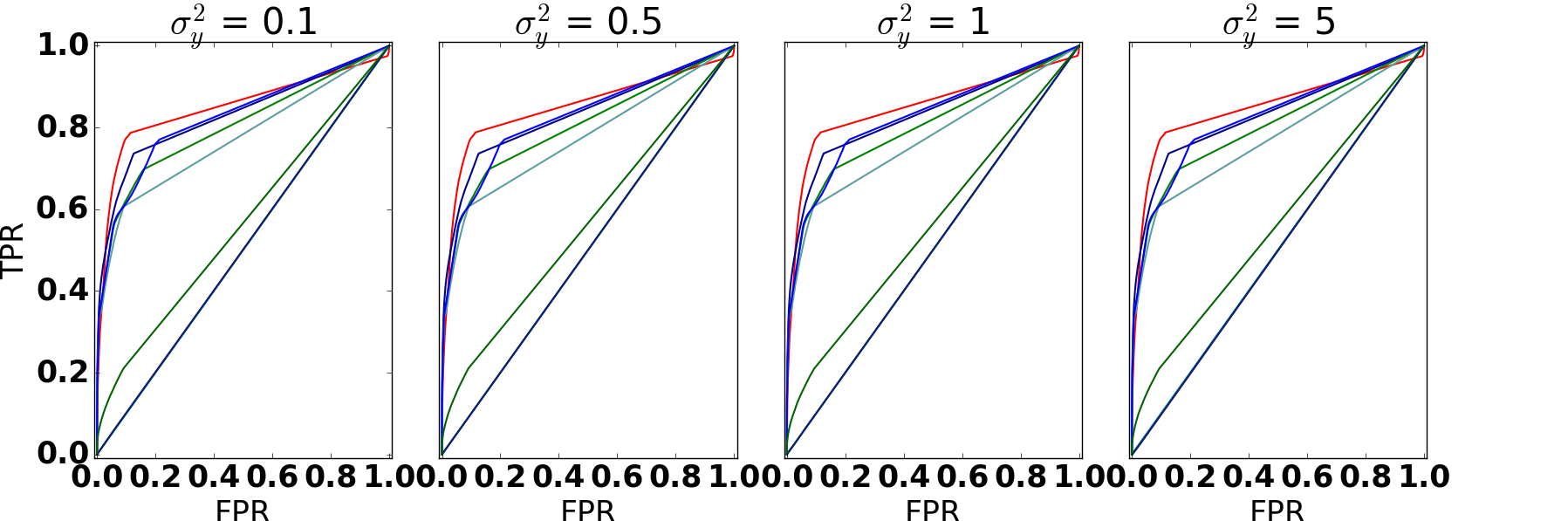}
    \label{excroc}}\hfill
\subfloat[Different magnitude of noise]{%
  \includegraphics[width=.48\textwidth]{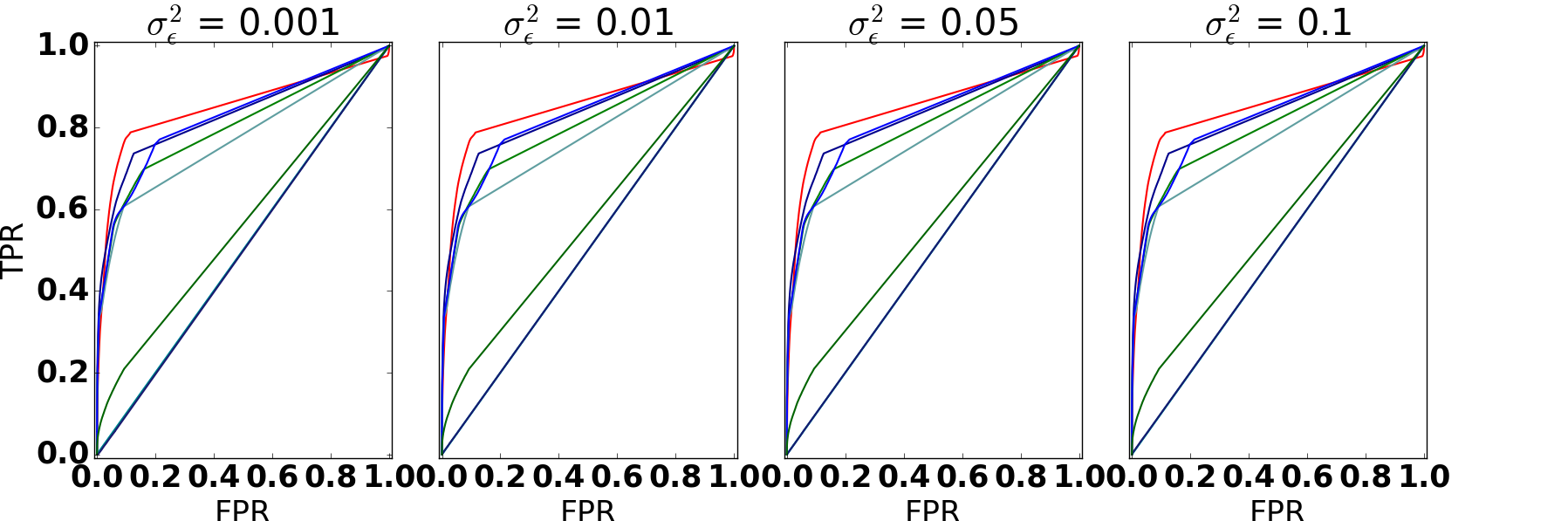}
  \label{exweroc}
}\\
  
\caption{ROC curves for experiments with different parameters.}
\label{fig:roc}
\end{figure*}

The Figure~\ref{fig:roc} shows the remaining images of ROC curves in synthetic experiments to compare our method with other existing methods by using the same parameters in our paper. For each configuration, the reported curve is drawn over five random seeds. And we can see that TgSLMM behaves almost always best.

\section*{Acknowledgment}
The authors would like to thank Haohan Wang from University of Illinois Urbana-Champaign and Changpeng Lu from Rutgers University for early-stage efforts of this project. We also appreciate Zhou Fang for providing computational resource for part of the experiments.



\EOD

\end{document}